\documentclass[10pt,showpacs,twocolumn,nofootinbib,aps,prd]{revtex4-1}
\usepackage{CJK}
\usepackage{amsmath,amssymb,bm,mathrsfs}
\usepackage{mathtools,mathptmx,array,slashed}
\usepackage{graphicx,graphics,subfigure}
\usepackage{ctable,multirow}
\usepackage{tabularx}
\usepackage[colorlinks=true,linkcolor=blue,citecolor=blue,urlcolor=blue]{hyperref}

\def\be{\begin{equation}}
\def\ee{\end{equation}}
\def\bea{\begin{eqnarray}}
\def\eea{\end{eqnarray}}
\def\ba{\begin{aligned}}
\def\ea{\end{aligned}}
\def\nn{\nonumber}
\def\p{\partial}

\begin{document}
\begin{CJK*}{GBK}{song}

\title{Topological classes of thermodynamics of rotating AdS black holes}

\author{Di Wu}
\email{wdcwnu@163.com}

\author{Shuang-Qing Wu}
\email{sqwu@cwnu.edu.cn}

\affiliation{School of Physics and Astronomy, China West Normal University, Nanchong,
Sichuan 637002, People's Republic of China}

\date{\today}

\begin{abstract}
In this paper, we extend our previous work [\href{http://dx.doi.org/10.1103/PhysRevD.107.024024}
{Phys. Rev. D \textbf{107}, 024024 (2023)}] to the more general cases with a negative cosmological
constant, and investigate the topological numbers for the singly rotating Kerr-AdS black holes in
all dimensions and the four-dimensional Kerr-Newman-AdS black hole as well as the three-dimensional
Ba\~nados-Teitelboim-Zanelli black hole. We find that the topological numbers of black holes are
remarkably influenced by the cosmological constant. In addition, we also demonstrate that the
dimension of spacetimes has an important effect on the topological number for rotating AdS black
holes. Furthermore, it is interesting to observe that the difference between the topological number
of the AdS black hole and that of its corresponding asymptotically flat black hole is always unity.
This new observation leads us to conjure that it might be valid also for other black holes. Of
course, this novel conjecture needs to be further verified by examining the topological numbers
of many other black holes and their AdS counterparts in the future work.
\end{abstract}

\maketitle
\end{CJK*}

\section{Introduction}

Recently, topology, as an important mathematical tool that applies to black hole physics, has
received considerable interest and enthusiasm. The current research on topology is mainly embodied
in two aspects. On the one hand, there is the research on the light rings \cite{PRL119-251102,
PRL124-181101,PRD102-064039,PRD103-104031} of some black holes, which may provide more footprints
for the observation of black holes and has been extended to timelike circular orbits \cite{2207.08397,
2301.04786}; on the other hand, there is the research on the thermodynamic topological classification
of various black holes \cite{PRL129-191101,PRD105-104003,PRD105-104053,PRD107-046013,2207.10612,
2211.12957,2302.01100,PRD106-064059,2208.10177,2211.05524,2211.15534,2212.04341,2302.03980,
PRD107-024024}.

In particular, a new method to investigate the thermodynamic topological properties of black
holes is proposed in Ref. \cite{PRL129-191101} by considering black hole solutions as topological
thermodynamic defects and constructing topological numbers, and further, dividing all black
holes into three categories according to their different topological numbers. Because these
topological numbers are universal constants that are independent of the black hole solution
parameters, hence they are very important for understanding the nature of black holes and gravity.
The topological approach proposed in Ref. \cite{PRL129-191101} quickly gained popularity due to
its straightforwardness and adaptability, and subsequently it was effectively used to explore
the topological numbers of several well-known black hole solutions \cite{PRD106-064059,2208.10177,
2211.05524,2211.15534,2212.04341,2302.03980}, i.e., the Schwarzschild-AdS black hole \cite{PRD106-064059},
the static black holes in Lovelock gravity \cite{2208.10177}, the static Gauss-Bonnet-AdS black
holes \cite{2211.05524}, the static black hole in nonlinear electrodynamics \cite{2211.15534},
and the static Born-Infeld AdS black hole \cite{2212.04341}, \textcolor{black}{as well as
some static hairy black holes \cite{2302.03980}}. However, all of the preceding researches
\cite{PRD106-064059,2208.10177,2211.05524,2211.15534,2212.04341,2302.03980} are limited to the static
cases, leaving the topological numbers of rotating black holes \textcolor{black}{and AdS scalar
hairy black holes} unexplored. Very recently, we have extended the topological approach to
rotating black hole cases and investigated the topological numbers for the cases of rotating
Kerr and Kerr-Newman black holes \cite{PRD107-024024}.

Since the study of the topological number of black holes is still in its infancy and the
topological number of the rotating AdS black holes remains virgin territory, it deserves to
be explored deeply. On the other hand, the study of rotating AdS black holes has already shed
light on the nature of gravity through gauge-gravity dualities \cite{ATMP2-231,PLB428-105,
ATMP2-253}, so it is very important and remarkable to investigate the topological number of
rotating AdS black holes. These two aspects motivate us to conduct the present work. In this
paper, we shall investigate the topological number of the $d$-dimensional singly rotating
Kerr-AdS black holes and the four-dimensional Kerr-Newman-AdS black hole, as well as the
three-dimensional Ba\~nados-Teitelboim-Zanelli (BTZ) black hole \cite{PRL69-1849,PRD48-1506,
CQG12-2853}. Compared with the previous paper \cite{PRD107-024024}, the aim of the present
work is concentrated on investigating the impact of the cosmological constant on the topological
number of black holes, which has not been studied in any previous related literature. We will
see that the cosmological constant is important in determining the topological number of
rotating black holes, and observe that the difference between the topological number of
the AdS black hole and that of its corresponding asymptotically flat black hole is always
unity, which leads us to conjure that it might also hold true for other black holes.

The remaining part of this paper is organized as follows. In Sec. \ref{II}, we first give
a brief review of the topological approach and investigate the topological number of the
four-dimensional Schwarzschild-AdS black hole as a warmup exercise. In Sec. \ref{III}, we
will focus on the topological number of the four-dimensional rotating Kerr-AdS black hole.
In Sec. \ref{IV}, we shall extend these discussions to the cases of the $d$-dimensional
singly rotating Kerr-AdS black holes. In Sec. \ref{V}, we will investigate the topological
number of the three-dimensional rotating BTZ black hole. In Sec. \ref{VI}, we then turn to
discuss the topological number of the four-dimensional Kerr-Newman-AdS black hole. Finally,
we present our conclusions in Sec. \ref{VII}. In the Appendix \ref{App}, the topological
number of the three-dimensional charged BTZ black hole is also investigated.

\section{Schwarzschild-AdS$_4$ black hole}\label{II}

In this section, we first present a brief review of the topological approach proposed in
Ref. \cite{PRL129-191101}, then investigate the topological number of the four-dimensional
Schwarzschild-AdS black hole as a warmup exercise. There are two reasons for us to do so.
On the one hand, it can be used to show that the charge parameter has a significant effect
on the topological numbers of the static AdS$_4$ black holes. On the other hand, it is also
convenient for us to make a comparison with the corresponding results of the rotating
Kerr-AdS$_4$ black hole, so as to observe the influence of the rotation parameter on
the topological number of the Schwarzschild-AdS$_4$ black hole.

As shown in Ref. \cite{PRL129-191101}, one can introduce the generalized off-shell Helmholtz
free energy
\be\label{FE}
\mathcal{F} = M -\frac{S}{\tau} \, ,
\ee
for a black hole thermodynamic system with mass $M$ and entropy $S$, where $\tau$ is an extra
variable that can be thought of as the inverse temperature of the cavity enclosing the black
hole. Only when $\tau = 1/T$ does the generalized Helmholtz free energy become on-shell.

In Ref. \cite{PRL129-191101}, a core vector $\phi$ is defined as\footnote{One can also construct
the vector $\phi$ in a more general form as $$\phi = \Big(\frac{\p\mathcal{F}}{\p{}r_{h}}\, ,
~-C\cot\Theta\csc\Theta\Big)\, ,$$ where $C$ is an arbitrary positive constant. Changing $C$
to a different value will slightly change the direction of the unit vector $n$, but will not
change the position of the zero point of the vector field or its corresponding winding number.
Therefore, one can just set $C = 1$ for the sake of simplicity. \textcolor{black}{However, the
authors of a new preprint \cite{2302.06201} criticize that this definition of the vector field
$\phi$ in Ref. \cite{PRL129-191101} is not intrinsic.}}
\bea
\phi = \Big(\frac{\p \mathcal{F}}{\p r_{h}}\, , ~-\cot\Theta\csc\Theta\Big) \, ,
\eea
in which the two parameters $r_h$ and $\Theta$ obey $0 < r_h < +\infty$ and $0\le \Theta\le \pi$,
respectively. The component $\phi^\Theta$ is divergent at $\Theta = 0,\pi$, thus the direction of
the vector points outward there.

Using Duan's $\phi$-mapping topological current theory \cite{SS9-1072,NPB514-705,PRD61-045004},
a topological current can be described as follows:
\be\label{jmu}
j^{\mu}=\frac{1}{2\pi}\epsilon^{\mu\nu\rho}\epsilon_{ab}\p_{\nu}n^{a}\p_{\rho}n^{b}\, , \qquad
\mu,\nu,\rho=0,1,2,
\ee
where $\p_{\nu}= \p/\p{}x^{\nu}$ and $x^{\nu}=(\tau,~r_h,~\Theta)$. The unit vector $n$ reads as
$n = (n^r, n^\Theta)$, where $n^r = \phi^{r_h}/||\phi||$ and $n^\Theta = \phi^{\Theta}/||\phi||$.
It is simple to demonstrate that the topological current (\ref{jmu}) given above is conserved,
allowing one to easily deduce $\p_{\mu}j^{\mu} = 0$. It is then established that the topological
current $j^\mu$ is a $\delta$-function of the field configuration \cite{NPB514-705,PRD61-045004}:
\be
j^{\mu}=\delta^{2}(\phi)J^{\mu}\Big(\frac{\phi}{x}\Big)\, ,
\ee
where the 3-dimensional Jacobian $J^{\mu}\big(\phi/x\big)$ is defined as: $\epsilon^{ab}J^{\mu}
\big(\phi/x\big) = \epsilon^{\mu\nu\rho}\p_{\nu}\phi^a\p_{\rho}\phi^b$. It is simple to indicate
that $j^\mu$ equals to zero only when $\phi^a(x_i) = 0$, hence the topological number $W$ can
be determined as follows:
\be
W = \int_{\Sigma}j^{0}d^2x = \sum_{i=1}^{N}\beta_{i}\eta_{i} = \sum_{i=1}^{N}w_{i}\, ,
\ee
where $\beta_i$ is the positive Hopf index that counts the number of the loops of the vector
$\phi^a$ in the $\phi$-space when $x^{\mu}$ are around the zero point $z_i$, whilst $\eta_{i}=
\textit{sign}(J^{0}({\phi}/{x})_{z_i})=\pm 1$ is the Brouwer degree, and $w_{i}$ is the winding
number for the $i$-th zero point of $\phi$ that is contained in $\Sigma$. It is important to
keep in mind that two loops $\Sigma_1$ and $\Sigma_2$ have the same winding number if they
both enclose the same zero point of $\phi$. On the other hand, if there is no zero point in
the surrounding region, then one can arrive at the topological number: $W = 0$.

In the following, we shall investigate the topological number of the four-dimensional
Schwarzschild-AdS black hole via the above topological approach. For the Schwarzschild-AdS$_4$
black hole, its metric has the form \cite{CMP10-280}
\bea\label{4dSchAdS}
ds^2 &=& -f(r)dt^2 +\frac{dr^2}{f(r)} +r^2\big(d\theta^2 +\sin^2\theta{}d\varphi^2\big) \, ,
\eea
where
\be
f(r) = 1 -\frac{2m}{r} +\frac{r^2}{l^2} \, , \nn
\ee
in which $m$ is the mass parameter, and $l$ is the cosmological scale associated with the
pressure $P = 3/(8\pi{}l^2)$ of the four-dimensional AdS black holes \cite{CPL23-1096,
CQG26-195011,PRD84-024037}. The mass and entropy associated with the above solution
(\ref{4dSchAdS}) can be computed via the standard method and have the following exquisite
forms:
\be
M = m \, , \qquad{} S = \pi{} r_h^2 \, ,
\ee
where $r_h$ are the locations of the event and Cauchy horizons that satisfy the equation:
$f(r_h) = 0$.

For the Schwarzschild-AdS$_4$ black hole, one can define the generalized Helmholtz free
energy as
\be
\mathcal{F} = \frac{r_h}{2} +\frac{4\pi}{3}Pr_h^2 -\frac{\pi{} r_h}{\tau} \, .
\ee
The components of the vector $\phi$ can be easily calculated as:
\be
\phi^{r_h} = \frac{1}{2} +4\pi{} Pr_h^2 -\frac{2\pi{} r_h}{\tau} \, , \quad
\phi^{\Theta} = -\cot\Theta\csc\Theta \, .
\ee
By solving the equation $\phi^{r_h} = 0$, one can obtain a curve on the $r_h-\tau$ plane.
For the four-dimensional Schwarzschild-AdS black hole, one can get
\be
\tau = \frac{4\pi{} r_h}{1 +8\pi{} Pr_h^2} \, .
\ee

\begin{figure}[t]
\centering
\includegraphics[width=0.4\textwidth]{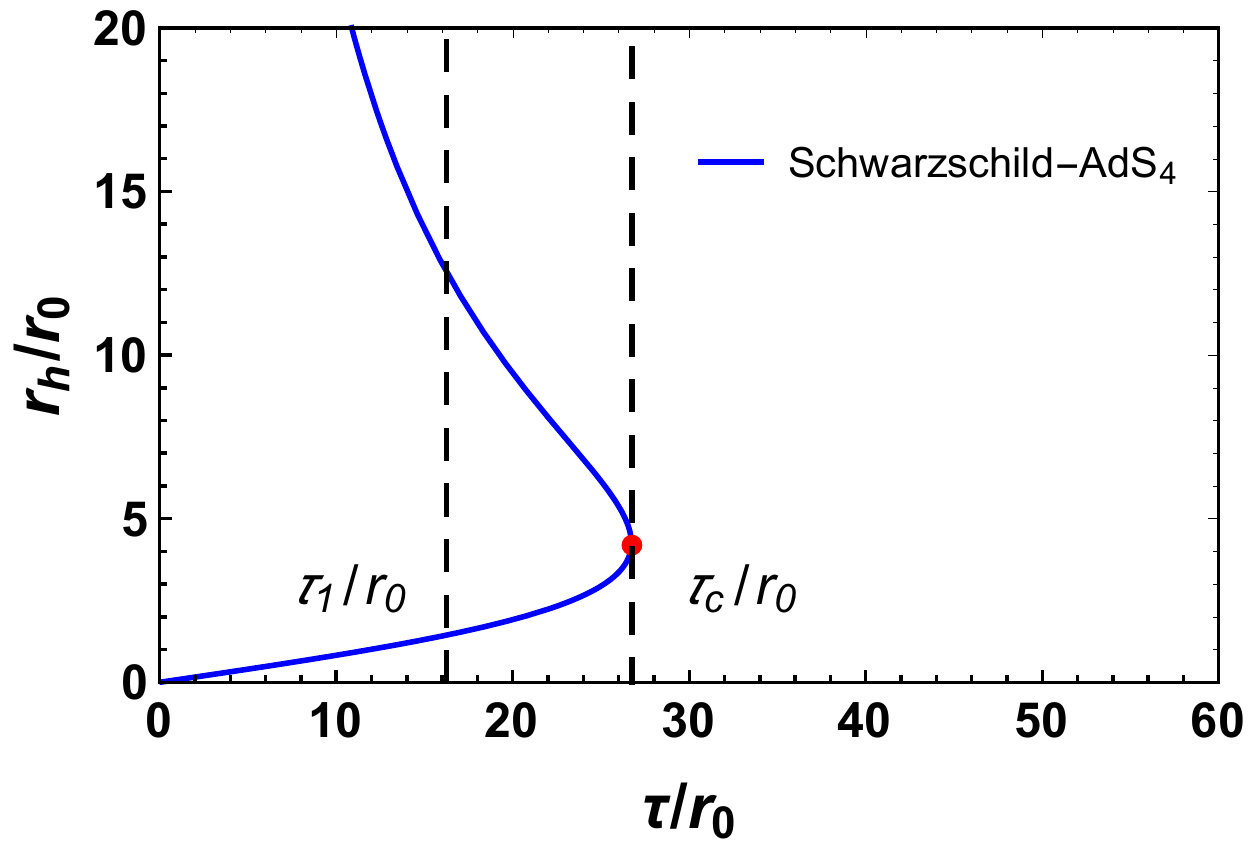}
\caption{Zero points of the vector $\phi^{r_h}$ shown on the $r_h-\tau$ plane
with $Pr_0^2=0.0022$ for the Schwarzschild-AdS$_4$ black hole. The annihilation
point for this black hole is represented by the red dot with $\tau_c$. There
are two Schwarzschild-AdS$_4$ black holes when $\tau = \tau_1$. Obviously, the
topological number is: $W = 1-1 = 0$. \label{SchAdS}}
\end{figure}

Taking the pressure $Pr_0^2 = 0.0022$, where $r_0$ is an arbitrary length scale set by the
size of a cavity enclosing the black hole, we show zero points of $\phi^{r_h}$ in the $r_h-\tau$
plane in Fig. \ref{SchAdS}. For small $\tau$, such as $\tau = \tau_1$, there are two intersection
points for the Schwarzschild-AdS$_4$ black hole. The intersection points exactly satisfy the
condition $\tau = 1/T$, and thus represent the on-shell Schwarzschild-AdS$_4$ black holes with the
characteristic temperature $T = 1/\tau$. The two intersection points for the Schwarzschild-AdS$_4$
black hole can coincide with each other when $\tau = \tau_c$, and then vanish when $\tau > \tau_c$,
therefore $\tau_c$ is an annihilation point and can be found at $\tau_c = 26.72r_0$, which can be
seen straightforwardly from Fig. \ref{SchAdS}. Furthermore, the annihilation point $\tau_c$ divides
the Schwarzschild-AdS$_4$ black hole into the upper and lower branches with the winding numbers
$w = 1$ and $w = -1$, respectively. One can see that, for the Schwarzschild-AdS$_4$ solutions
at any given temperature, there can exist one thermodynamically stable black hole and one
thermodynamically unstable black hole. It has been shown that the winding number can be used
to characterize the local thermodynamic stability, with positive and negative values corresponding
to thermodynamically stable and unstable black holes \cite{PRL129-191101}, respectively.

\begin{figure}[h]
\centering
\includegraphics[width=0.35\textwidth]{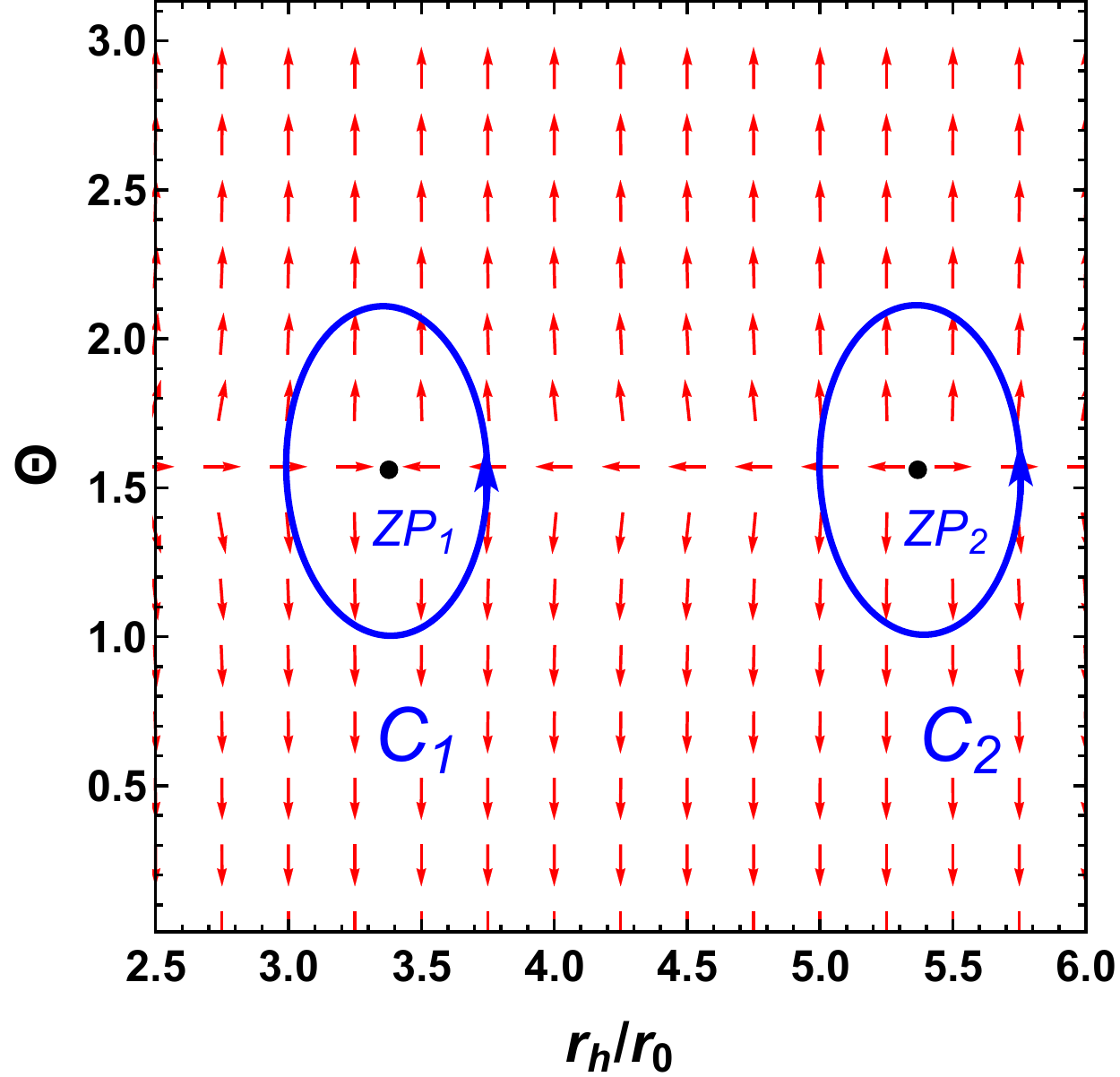}
\caption{The red arrows represent the unit vector field $n$ on a portion of the $r_h-\Theta$
plane with $Pr_0^2=0.0022$ and $\tau/r_0=26$ for the Schwarzschild-AdS$_4$ black hole. The
zero points (ZPs) marked with black dots are at $(r_h/r_0, \Theta) = (3.36,\pi/2)$, $(5.38,
\pi/2)$ for ZP$_1$ and ZP$_2$, respectively. The blue contours $C_i$ are closed loops
surrounding the zero points. \label{SchAdS4d}}
\end{figure}

Alternatively, the unit vector field $n$ can also be plotted for any arbitrarily selected
typical values (keep in mind that $\tau$ must be less than $\tau_c$), for instance, $\tau/r_0
= 26$ and $Pr_0^2 = 0.0022$ in Fig. \ref{SchAdS4d}, where we find two zero points: ZP$_1$ at
$r_h = 3.36r_0$ and ZP$_2$ at $r_h = 5.38r_0$, with the winding numbers $w_1 = 1$ and $w_2 = -1$,
respectively, to determine the topological number for the four-dimensional Schwarzschild-AdS
black hole. Based upon the local property of the zero point, one can easily find that the
topological number is: $W = -1 +1 = 0$ for the Schwarzschild-AdS$_4$ black hole, which is
consistent with the result given in Ref. \cite{PRD106-064059}.

In addition, the fact that the topological number of the Schwarzschild-AdS$_4$ black hole is
zero while that of the Schwarzschild black hole is $-1$ \cite{PRL129-191101} suggests that
the cosmological constant significantly changes the topological number of the static black
holes. Furthermore, since the topological number of the RN-AdS$_4$ black hole is: $W = 1$,
it is easy to see that the Schwarzschild-AdS$_4$ black hole and the RN-AdS$_4$ black hole
belong to two different topological classes according to the topological classification
method proposed in Ref. \cite{PRL129-191101}, which indicates that the electric charge has
an important influence on the topological number of static AdS$_4$ black holes.

\section{Kerr-AdS$_4$ black hole}\label{III}

From now on, we come to the main subject of this paper, i.e., exploring the topological number
of rotating AdS black holes. In this section, we will focus on the topological number of the
four-dimensional Kerr-AdS black hole, whose metric in the asymptotically non-rotating frame
has the form \cite{CMP10-280}
\bea\label{KerrAdS}
ds^2 &=& -\frac{\Delta_r}{\Sigma}\Big(\frac{\Delta_\theta}{\Xi}dt
 -\frac{a}{\Xi}\sin^2\theta{} d\varphi \Big)^2
 +\frac{\Sigma}{\Delta_r}dr^2 +\frac{\Sigma}{\Delta_\theta}d\theta^2 \nn \\
&&+\frac{\Delta_\theta\sin^2\theta}{\Sigma}\Big[\frac{a(r^2 +l^2)}{l^2\Xi}dt
 -\frac{r^2 +a^2}{\Xi}d\varphi \Big]^2 \,
\eea
in terms of the Boyer-Lindquist coordinates, where
\bea
&&\Delta_r = (r^2 +a^2)\Big(1 +\frac{r^2}{l^2}\Big) -2mr \, , \quad
\Xi = 1 -\frac{a^2}{l^2} \, , \nn \\
&&\Delta_\theta = 1 -\frac{a^2}{l^2}\cos^2\theta \, , \quad
\Sigma = r^2 +a^2\cos^2\theta \, , \nn
\eea
in which $a$ is the rotation parameter, $m$ is the mass parameter, and $l$ is the AdS radius.

The mass $M$ and entropy $S$ associated with the above solution (\ref{KerrAdS}) are
\cite{CQG17-399}
\be
M = \frac{m}{\Xi^2} \, , \qquad  S = \frac{\pi(r_h^2 +a^2)}{\Xi} \, .
\ee
Using the definition of the generalized off-shell Helmholtz free energy (\ref{FE}) and
$l^2 = 3/(8\pi{}P)$, one can easily get
\be
\mathcal{F} = \frac{3(r_h^2 +a^2)\big[2\pi{}r_h(8\pi{} Pa^2 +4Pr_h\tau -3)
 +3\tau\big]}{2r_h(8\pi{}Pa^2 -3)^2\tau}
\ee
for the Kerr-AdS$_4$ black hole. Then the components of the vector $\phi$ can be computed as
\bea
\phi^{r_h} &=& \frac{12\pi(8\pi{}Pa^2 -3)r_h^3 +3a^2(8\pi{}Pr_h^2 -3)\tau}{2r_h^2
 (8\pi{}Pa^2 -3)^2\tau} \nn \\
&& +\frac{9(1 +8\pi{}Pr_h^2)}{2(8\pi{}Pa^2 -3)^2} \, , \\
\phi^{\Theta} &=& -\cot\Theta\csc\Theta \, .
\eea
By solving the equation $\phi^{r_h} = 0$, one can obtain
\be
\tau = \frac{4\pi{}r_h^3(3 -8\pi{}Pa^2)}{a^2(8\pi{}Pr_h^2 -3) +3(8\pi{}Pr_h^2 +1)r_h^2}
\ee
as the zero point of the vector field $\phi$.

\begin{figure}[h]
\centering
\includegraphics[width=0.4\textwidth]{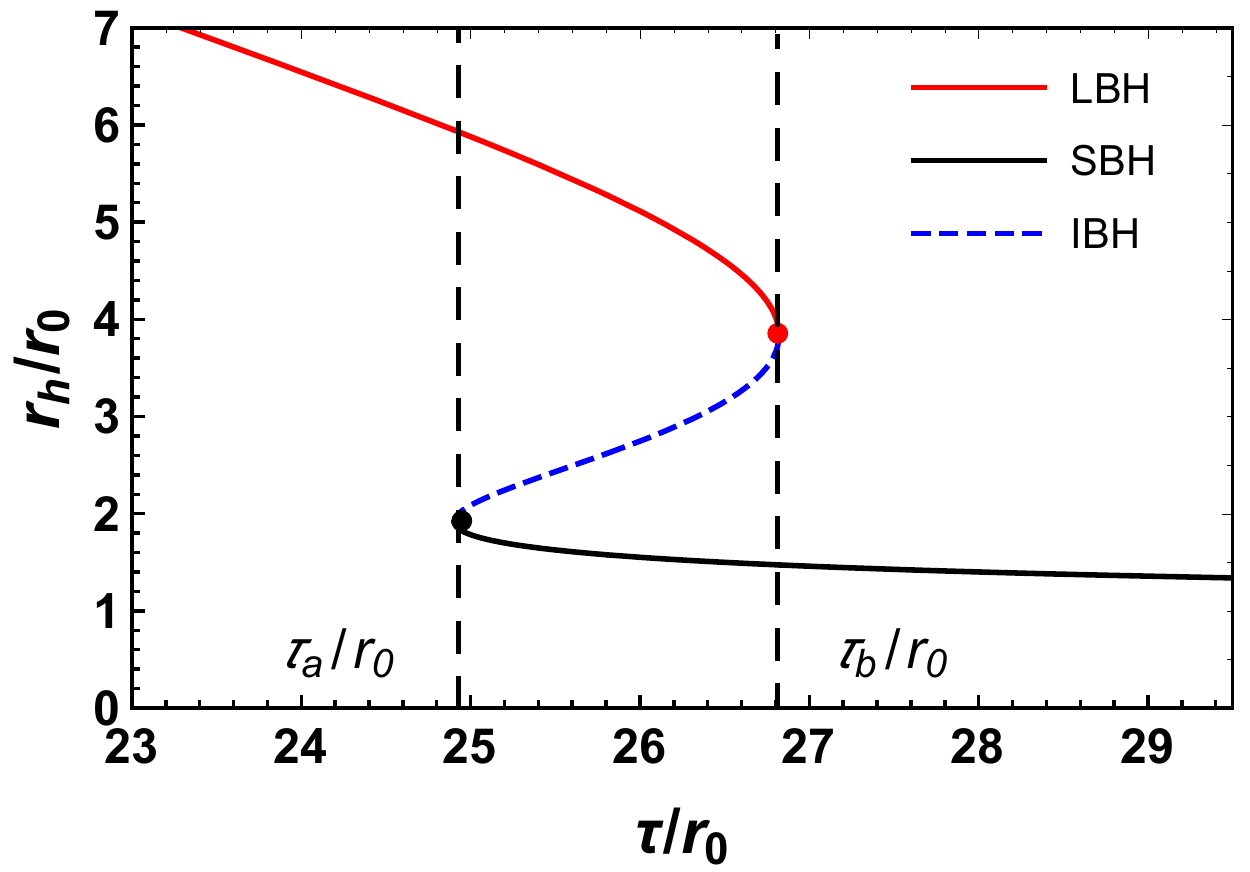}
\caption{Zero points of $\phi^{r_h}$ shown in the $r_h-\tau$ plane  with $Pr_0^2=0.0022$
and $a=r_0$ for the Kerr-AdS$_4$ black hole. The red solid, blue dashed, and black solid
lines are for the large black hole (LBH), intermediate black hole (IBH), and small black
hole (SBH), respectively. The annihilation and generation points are represented by red
and black dots, respectively. \label{4dKerrAdS}}
\end{figure}

\begin{figure}
\subfigure[~{The unit vector field for the Kerr-AdS$_4$ black hole with
$\tau/r_0=24$, $a/r_0=1$, and $Pr_0^2=0.022$.}]{\label{4dKerrAdStau24}
\includegraphics[width=0.35\textwidth]{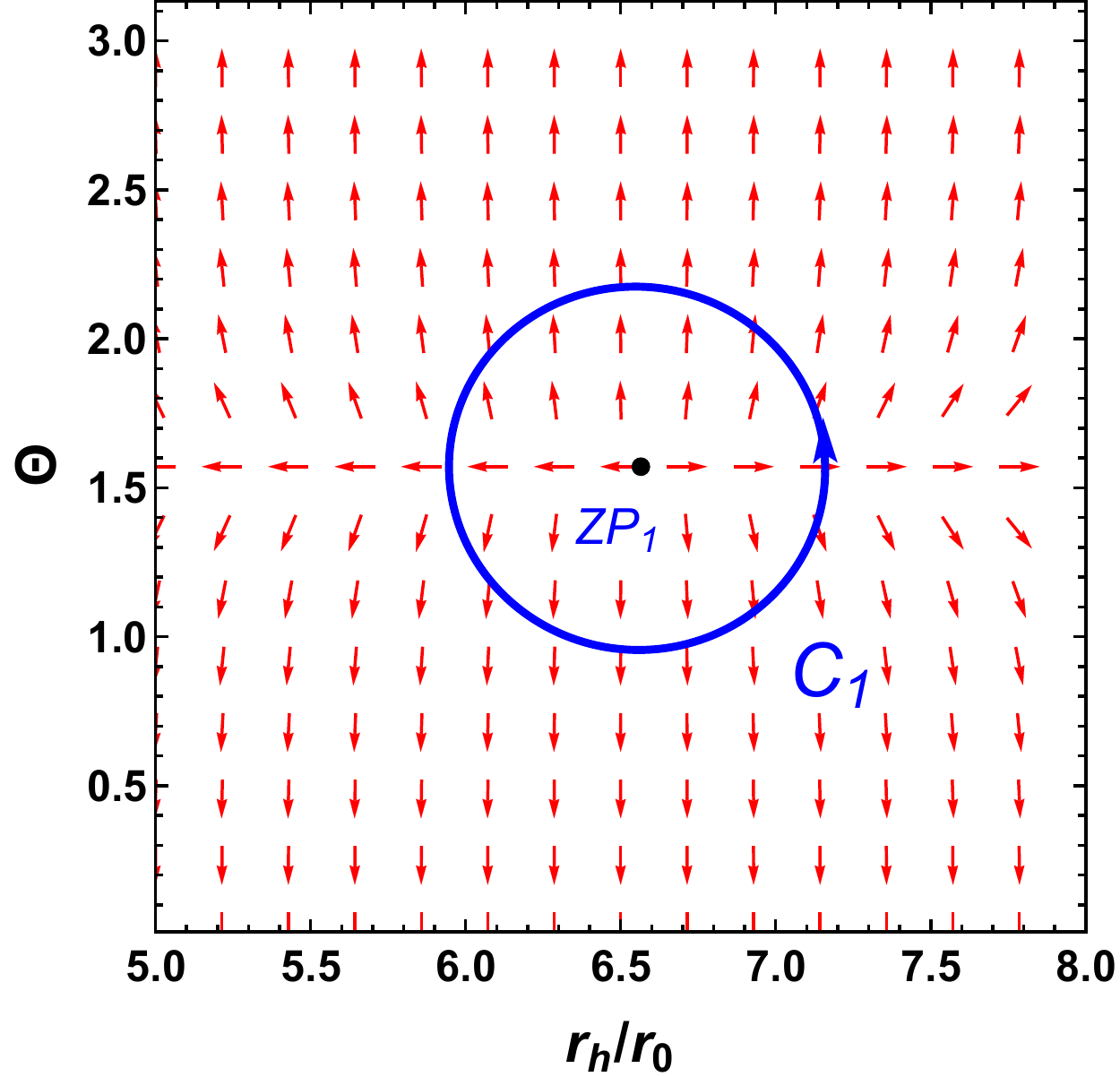}}
\subfigure[~{The unit vector field for the Kerr-AdS$_4$ black hole with
$\tau/r_0=26$, $a/r_0=1$, and $Pr_0^2=0.022$.}]{\label{4dKerrAdStau26}
\includegraphics[width=0.35\textwidth]{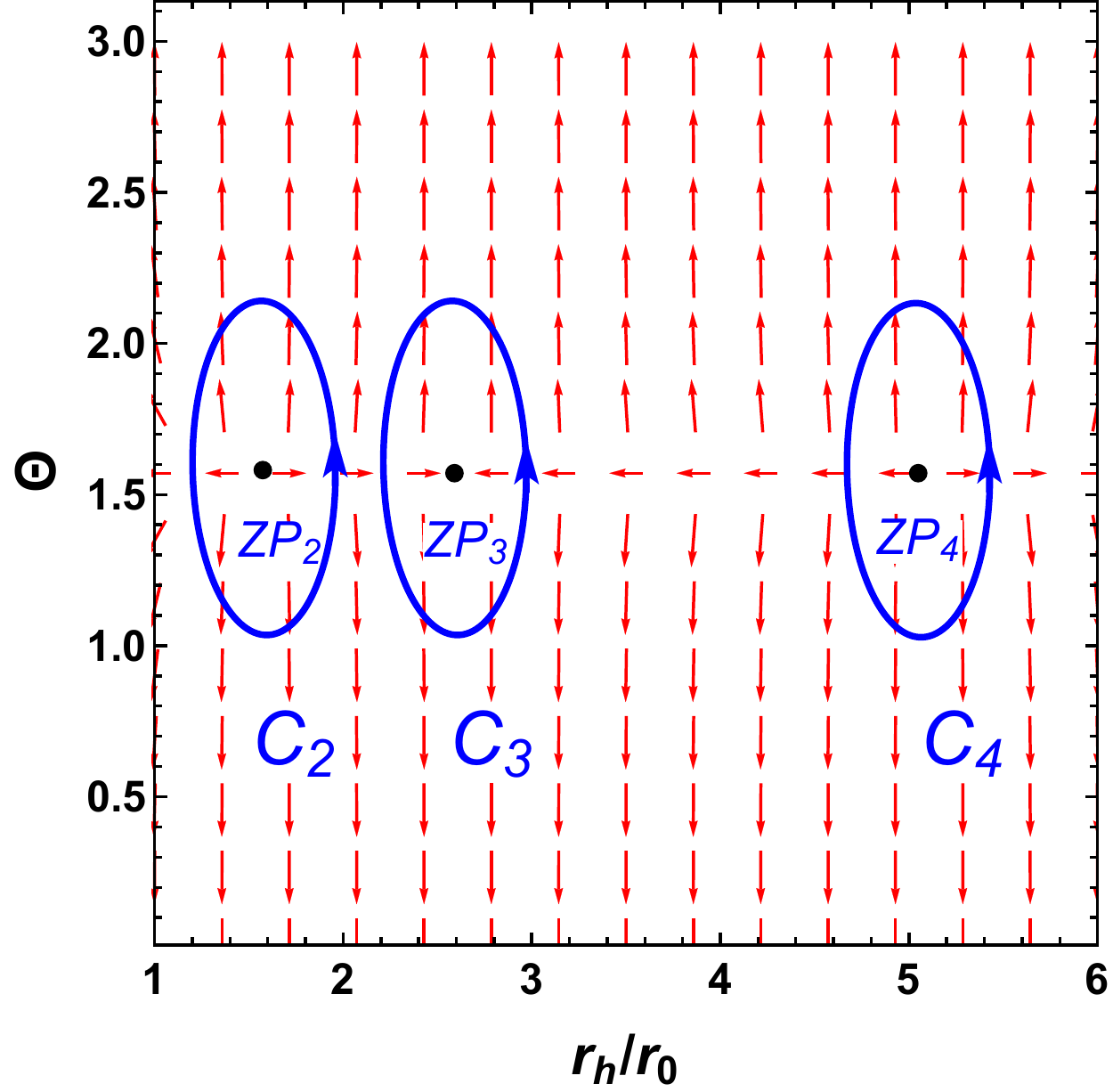}}
\subfigure[~{The unit vector field for the Kerr-AdS$_4$ black hole with
$\tau/r_0=28$, $a/r_0=1$, and $Pr_0^2=0.022$.}]{\label{4dKerrAdStau28}
\includegraphics[width=0.35\textwidth]{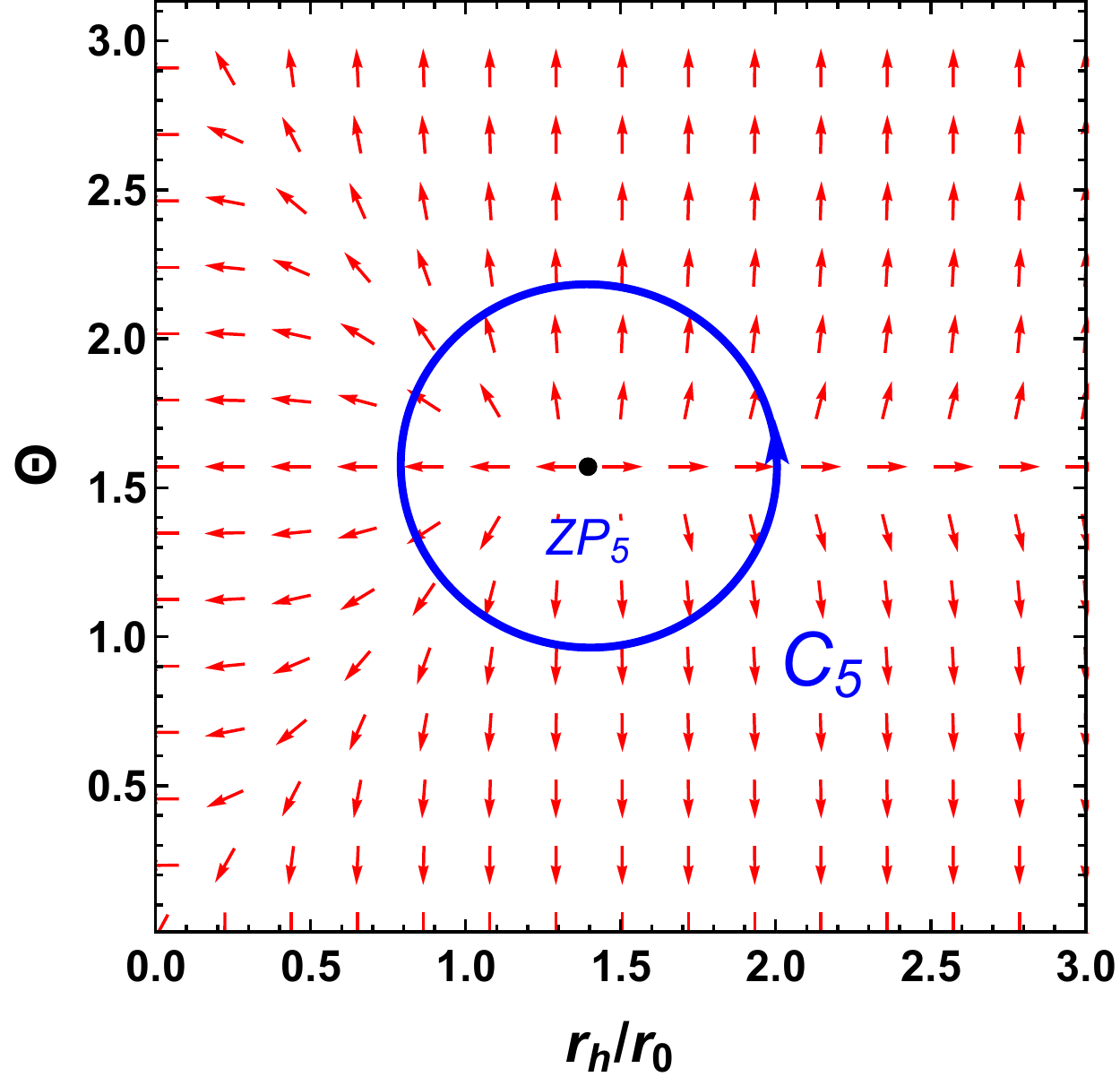}}
\caption{The red arrows represent the unit vector field $n$ on a portion of the $r_h-\Theta$
plane. The zero points (ZPs) marked with black dots are at $(r_h/r_0,\Theta) = (6.55,\pi/2)$,
$(1.55,\pi/2)$, $(2.75,\pi/2)$, $(5.11,\pi/2)$, $(1.40,\pi/2)$, for ZP$_1$, ZP$_2$, ZP$_3$,
ZP$_4$, and ZP$_5$, respectively. The blue contours $C_i$ are closed loops surrounding the
zero points. \label{KerrAdS4d}}
\end{figure}

Taking the pressure $Pr_0^2 = 0.0022$ and the rotation parameter $a = r_0$ for the Kerr-AdS$_4$
black hole, we show zero points of $\phi^{r_h}$ in the $r_h-\tau$ plane in Fig. \ref{4dKerrAdS},
and the unit vector field $n$ in Fig. \ref{KerrAdS4d} with $\tau = 24r_0$, $26r_0$, and $28r_0$,
respectively. From Figs. \ref{4dKerrAdS} and \ref{KerrAdS4d}, one can observe that for these
values of $Pr_0^2$ and $a/r_0$, one generation point and one annihilation point can be found at
$\tau/r_0 = \tau_a/r_0 = 24.90$ and $\tau/r_0 = \tau_b/r_0 = 26.82$, respectively. One can see
that there are one large black hole branch for $\tau < \tau_a$, three black hole branches for
$\tau_a < \tau < \tau_b$, and one small black hole branch for $\tau > \tau_b$. Calculating
the winding number $w$ for these three black hole branches, we find that both the small and
large black hole branches have $w = 1$, while the intermediate black hole branch has $w = -1$.
The Kerr-AdS$_4$ black hole always has the topological number $W = 1$, unlike the Kerr black hole,
which has a topological number of zero \cite{PRD107-024024}. Therefore, from the thermodynamic
topological standpoint, the Kerr-AdS$_4$ black hole and the Kerr black hole are different
kinds of black hole solutions, and this indicates that the cosmological constant is important
in determining the topological number for the rotating black hole. What is more, since the
topological number of the Schwarzschild-AdS$_4$ black hole is zero, while that of the Kerr-AdS$_4$
black hole is $1$, so it can be inferred that the rotation parameter has a remarkable effect on
the topological number for the uncharged AdS$_4$ black hole.

\section{Singly rotating Kerr-AdS black holes in arbitrary dimensions}\label{IV}

In this section, we will extend the above discussions to the cases of higher-dimensional
rotating black holes by considering the singly rotating Kerr-AdS solutions in arbitrary
dimensions. For $d$-dimensional singly rotating Kerr-AdS black holes, the metric in the
asymptotically non-rotating frame has the form \cite{PRD59-064005,PRL93-171102}
\bea\label{hdKerrAdS}
ds^2 &=& -\frac{\Delta_r}{\Sigma}\Big(\frac{\Delta_\theta}{\Xi}dt
 -\frac{a}{\Xi}\sin^2\theta{} d\varphi \Big)^2
 +\frac{\Sigma}{\Delta_r}dr^2 +\frac{\Sigma}{\Delta_\theta}d\theta^2 \nn \\
&&+\frac{\Delta_\theta\sin^2\theta}{\Sigma}\Big[\frac{a(r^2 +l^2)}{l^2\Xi}dt
 -\frac{r^2 +a^2}{\Xi}d\varphi \Big]^2 \nn \\
&&+r^2\cos^2\theta d\Omega_{d-4}^2 \, , \qquad
\eea
where $d\Omega_{d-4}$ denotes the line element of the ($d-4$)-dimensional unit sphere, and
\bea
&&\Delta_r = (r^2 +a^2)\Big(1 +\frac{r^2}{l^2}\Big) -2mr^{5-d} \, , \quad
\Xi = 1 -\frac{a^2}{l^2} \, , \nn \\
&&\Delta_\theta = 1 -\frac{a^2}{l^2}\cos^2\theta \, , \quad
\Sigma = r^2 +a^2\cos^2\theta \, . \nn
\eea

The thermodynamic quantities are \cite{PRD93-084015}
\be\ba\label{ThermhdKerrAdS}
&M = \frac{\omega_{d-2}m}{4\pi\Xi^2}\Big[\frac{(d-4)\Xi}{2} +1 \Big] \, ,
\quad J = \frac{\omega_{d-2}ma}{4\pi\Xi^2} \, , \\
&\Omega = \frac{a(r_h^2 +l^2)}{l^2(r_h^2 +a^2)} \, ,
\quad S = \frac{\mathcal{A}}{4} = \frac{\omega_{d-2}}{4\Xi}(r_h^2 +a^2)r_h^{d-4} \, , \\
&T = \frac{r_h}{2\pi}\Big(1 +\frac{r_h^2}{l^2} \Big)\Big(\frac{1}{r_h^2 +a^2}
+\frac{d-3}{2r_h^2} \Big) -\frac{1}{2\pi{}r_h} \, , \\
&V = \frac{r_h\mathcal{A}}{d-1}\Big[1 +\frac{a^2(r_h^2 +l^2)}{(d-2)\Xi{}l^2r_h^2} \Big] \, ,
\quad P = \frac{(d-1)(d-2)}{16\pi{}l^2} \, ,
\ea\ee
where $\omega_{d-2} = 2\pi^{(d-1)/2}/\Gamma[(d-1)/2]$, and $r_h$ is determined by the
horizon equation: $\Delta_r = 0$.

In our previous paper \cite{PRD107-024024}, we have uniformly considered the topological numbers
of $d$-dimensional singly rotating Kerr black holes. Here, we will extend that work to the cases
of $d$-dimensional singly rotating Kerr-AdS black holes. Since there is one more additional
thermodynamic quantity associated with the cosmological constant to be included, it will be
more convenient to separately consider the topological numbers of different dimensions.

\subsection{$d = 5$ case}\label{IVA}

We first consider $d = 5$ case. From Eq. (\ref{ThermhdKerrAdS}), one can obtain the expression
of the generalized Helmholtz free energy as

\bea
\mathcal{F} &=& -\frac{\pi(r_h^2 +a^2)}{8\tau(4\pi{}Pa^2 -3)^2}
\Big[12\pi{}r_h(3 -4\pi{}Pa^2) \nn \\
&&+\tau(4\pi{}Pr_h^2 +3)(4\pi{}Pa^2 -9) \Big] \, ,
\eea
so the components of the vector $\phi$ can be computed as
\bea
\phi^{r_h} &=& \frac{\pi}{4\tau(4\pi{}Pa^2 -3)^2}\Big\{6\pi(4\pi{}Pa^2 -3)(3r_h^2 +a^2) \nn \\
&&-r_h\tau(4\pi{}Pa^2 -9)\big[4\pi{}P(2r_h^2 +a^2) +3 \big] \Big\} \, , \\
\phi^{\Theta} &=& -\cot\Theta\csc\Theta \, .
\eea
It is simple to obtain
\be
\tau= \frac{6\pi(4\pi{}Pa^2 -3)(3r_h^2 +a^2)}{r_h(4\pi{}Pa^2 -9)[4\pi{}P(2r_h^2 +a^2) +3]}
\ee
as the zero point of the vector field $\phi$.

\begin{figure}[b]
\centering
\includegraphics[width=0.4\textwidth]{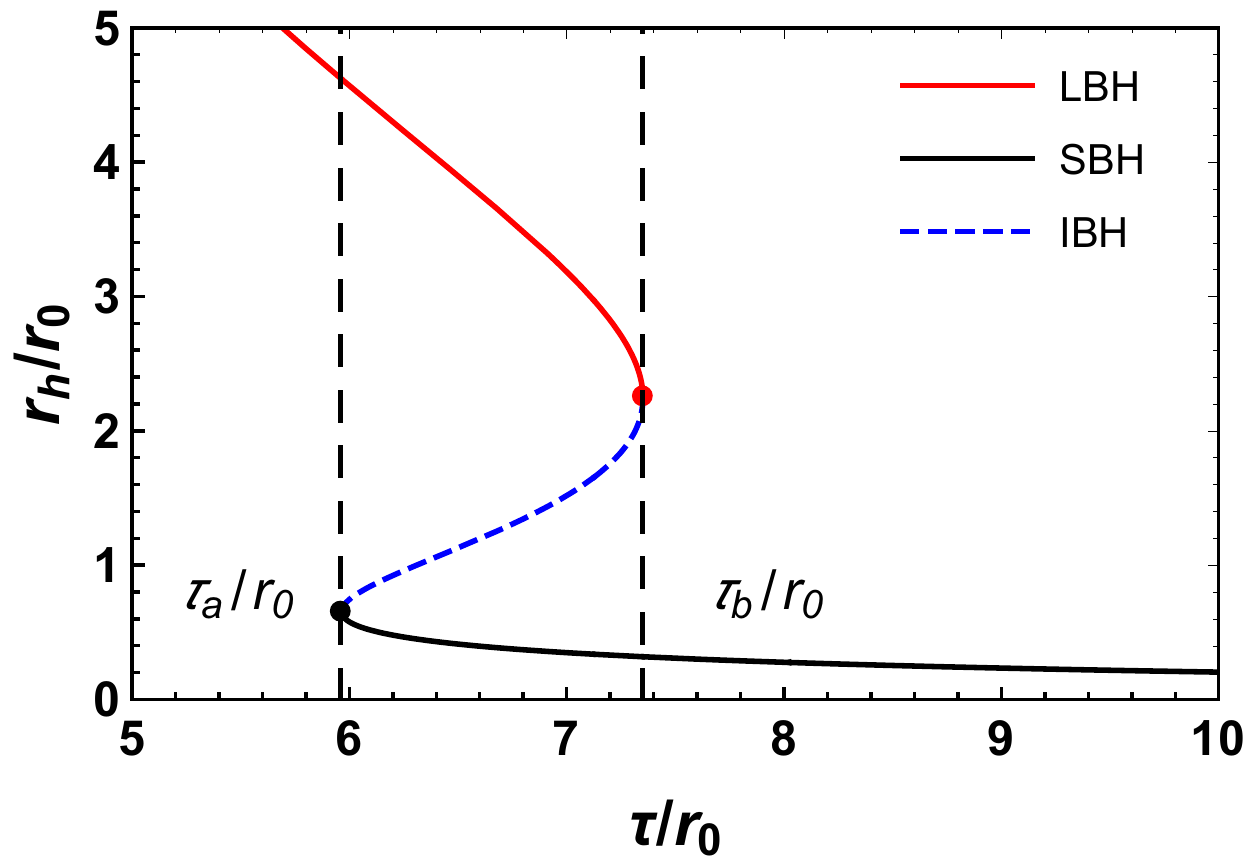}
\caption{The zero points of $\phi^{r_h}$ shown in the $r_h-\tau$ plane with
$Pr_0^2=0.02$ and $a/r_0=1$ for the singly rotating Kerr-AdS$_5$ black hole.
The red solid, blue dashed, and black solid lines are for the large black hole
(LBH), intermediate black hole (IBH), and small black hole (SBH), respectively.
The annihilation and generation points are represented by red and black dots,
respectively. \label{5dKerrAdS}}
\end{figure}

\begin{figure}{h}
\subfigure[~{The unit vector field for the five-dimensional singly
rotating Kerr-AdS black hole with $\tau/r_0=5$, $a/r_0=1$, and $Pr_0^2=0.02$.}]
{\label{5dKerrAdStau5}
\includegraphics[width=0.33\textwidth]{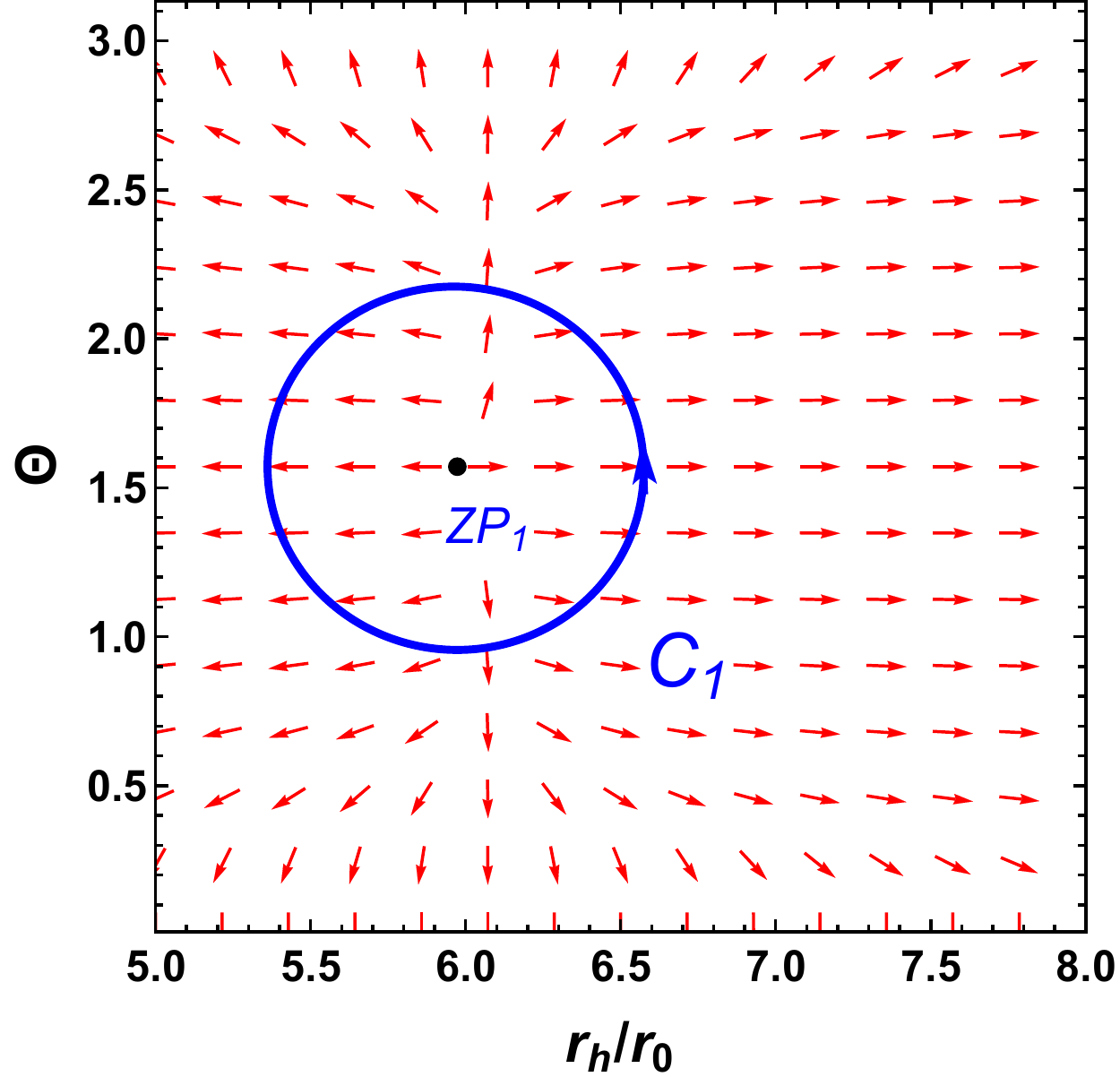}}
\subfigure[~{The unit vector field for the five-dimensional singly
rotating Kerr-AdS black hole with $\tau/r_0=7$, $a/r_0=1$, and $Pr_0^2=0.02$.}]
{\label{5dKerrAdStau7}
\includegraphics[width=0.33\textwidth]{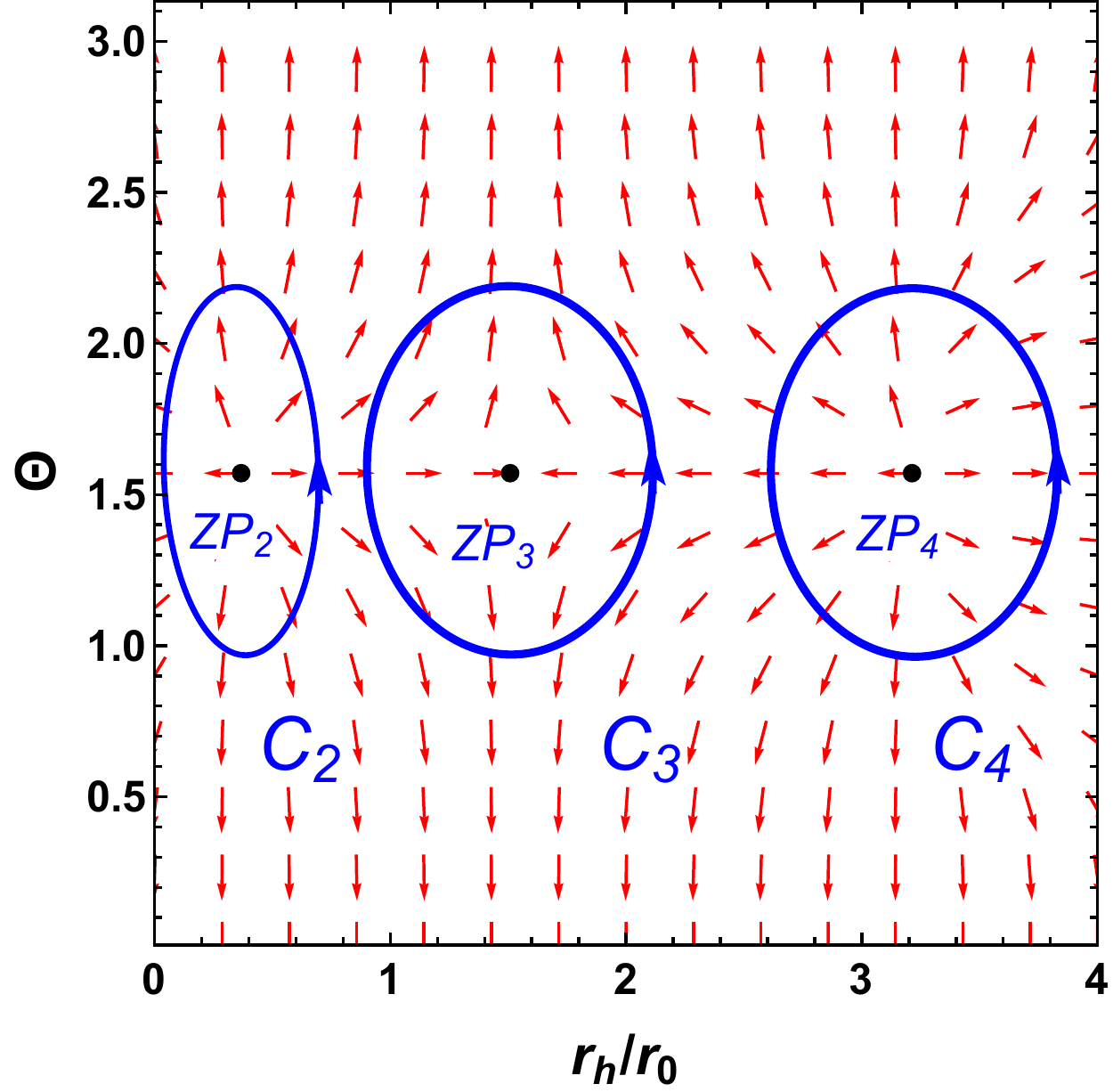}}
\subfigure[~{The unit vector field for the five-dimensional singly
rotating Kerr-AdS black hole with $\tau/r_0=9$, $a/r_0=1$, and $Pr_0^2=0.02$.}]
{\label{5dKerrAdStau9}
\includegraphics[width=0.33\textwidth]{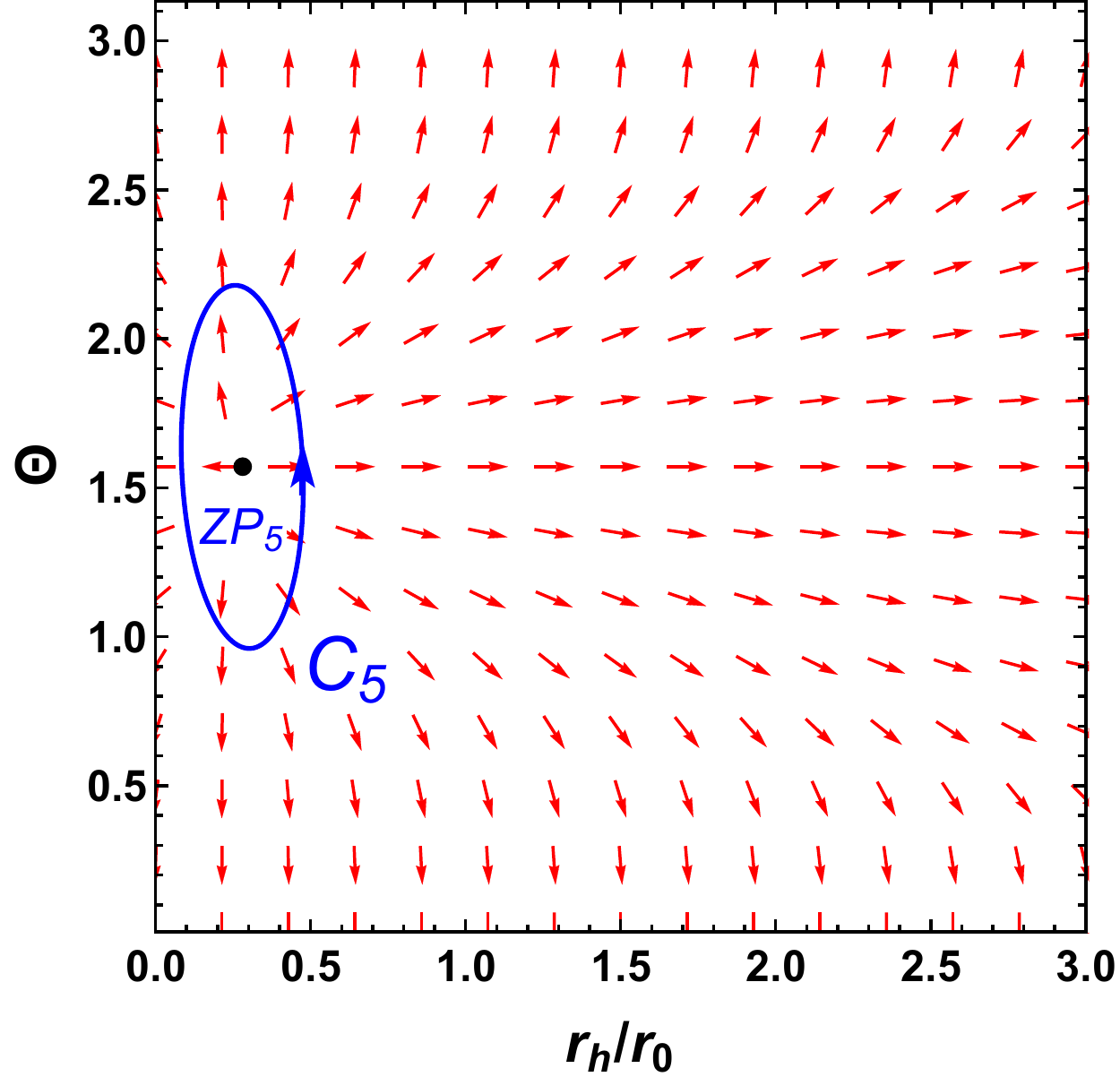}}
\caption{The red arrows represent the unit vector field $n$ on a portion of the $r_h-\Theta$
plane. The zero points (ZPs) marked with black dots are at $(r_h/r_0, \Theta) = (6.07,\pi/2)$,
$(0.35,\pi/2)$, $(1.52,\pi/2)$, $(3.18,\pi/2)$, $(0.23,\pi/2)$, for ZP$_1$, ZP$_2$, ZP$_3$,
ZP$_4$, and ZP$_5$, respectively. The blue contours $C_i$ are closed loops surrounding the
zero points. \label{KerrAdS5d}}
\end{figure}

For the singly rotating Kerr-AdS$_5$ black hole, we plot the zero points of the component
$\phi^{r_h}$  with $Pr_0^2 = 0.02$ and $a/r_0 = 1$ in Fig. \ref{5dKerrAdS}, and the unit
vector field $n$ in Fig. \ref{KerrAdS5d} with $\tau = 5r_0$, $7r_0$, and $9r_0$, respectively.
Note that for these values of $Pr_0^2$ and $a/r_0$, one generation point and one annihilation
point can be found in Fig. \ref{5dKerrAdS} at $\tau/r_0 = \tau_a/r_0 = 5.96$ and $\tau/r_0 =
\tau_b/r_0 = 7.35$, respectively. From Figs. \ref{5dKerrAdS} and \ref{KerrAdS5d}, one can easily
obtain the topological number $W = 1$ for the singly rotating Kerr-AdS$_5$ black hole using the
local property of the zero points, which is the same as that of the four-dimensional Kerr-AdS
black hole in the previous section but different from that of the five-dimensional singly
rotating Kerr black hole, which is $W = 0$ \cite{PRD107-024024}. Therefore, the topological
number of the five-dimensional rotating black hole is significantly changed when the cosmological
constant is turned on.

\subsection{$d = 6$ case}\label{IVB}

Next, we consider $d = 6$ case whose generalized Helmholtz free energy is
\bea
\mathcal{F} &=& -\frac{2\pi{}r_h(r_h^2 +a^2)}{3\tau(4\pi{}Pa^2 -5)^2}
 \Big[5\pi{}r_h(5 -4\pi{}Pa^2) \nn \\
&&+\tau(4\pi{}Pr_h^2 +5)(2\pi{}Pa^2 -5) \Big] \, .
\eea
Thus, the components of the vector $\phi$ are

\bea
\phi^{r_h} &=& \frac{2\pi}{3\tau(4\pi{}Pa^2 -5)}
 \Big\{10\pi{}r_h(4\pi{}Pa^2 -5)(2r_h^2 +a^2) \nn \\
&&-\tau(2\pi{}Pa^2 -5)\big[20\pi{}Pr_h^4 +3(4\pi{}Pa^2 +5)r_h^2 \nn \\
&&+5a^2 \big] \Big\} \, , \\
\phi^{\Theta} &=& -\cot\Theta\csc\Theta \, .
\eea
So the zero point of the vector field $\phi$ is
\be
\tau = \frac{10\pi{}r_h(4\pi{}Pa^2 -5)(2r_h^2 +a^2)}{(2\pi{}Pa^2 -5)
\big[20\pi{}Pr_h^4 +3(4\pi{}Pa^2 +5)r_h^2 +5a^2 \big]} \, .
\ee

\begin{figure}[h]
\centering
\includegraphics[width=0.4\textwidth]{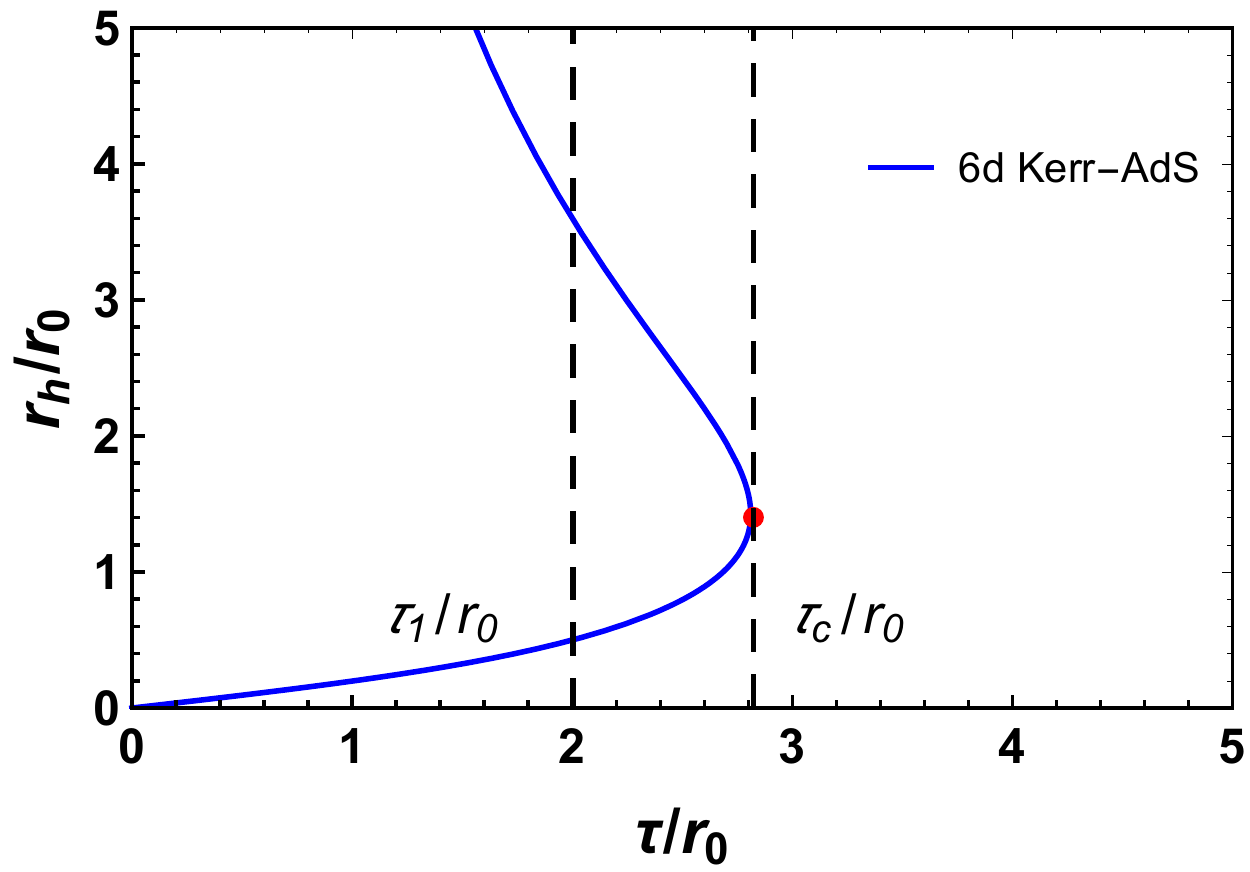}
\caption{Zero points of $\phi^{r_h}$ shown in the $r_h-\tau$ plane with $Pr_0^2=0.1$
and $a/r_0=1$ for the singly rotating Kerr-AdS$_6$ black hole. The red dot with
$\tau_c$ represents the annihilation point for the black hole. There are two singly
rotating Kerr-AdS$_6$ black holes when $\tau = \tau_1$. It is easy to obtain the
topological number: $W = 1-1 = 0$. \label{6dKerrAdS}}
\end{figure}

\begin{figure}[h]
\centering
\includegraphics[width=0.35\textwidth]{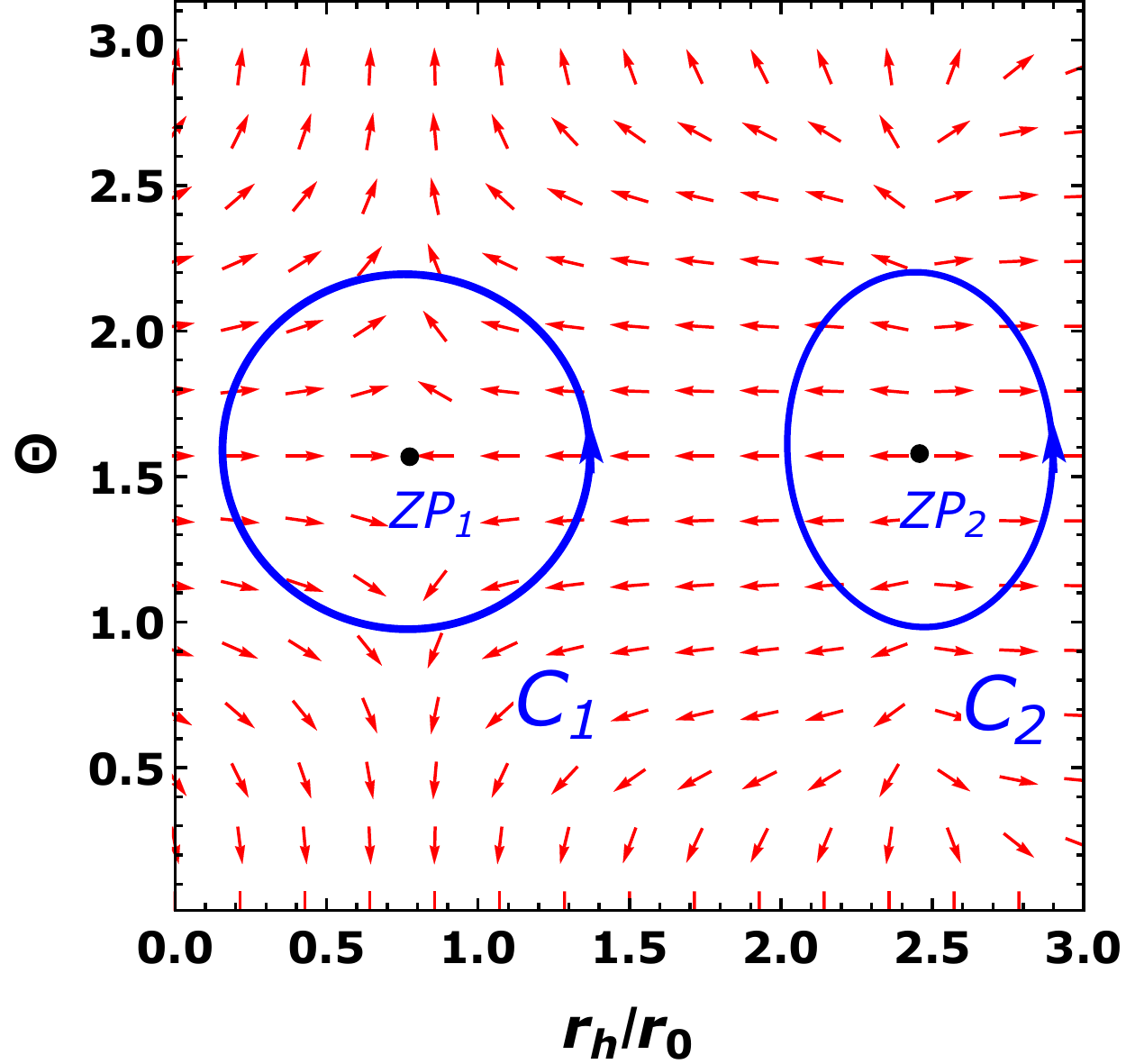}
\caption{The red arrows represent the unit vector field $n$ on a portion of the $r_h-\Theta$
plane with $Pr_0^2=0.1$, $a/r_0=1$, and $\tau/r_0=2.5$ for the singly rotating Kerr-AdS$_6$
black hole. The zero points (ZPs) marked with black dots are at $(r_h/r_0, \Theta)=(0.80,
\pi/2)$, $(2.43,\pi/2)$ for ZP$_1$ and ZP$_2$, respectively. The blue contours $C_i$ are
closed loops surrounding the zero points. \label{KerrAdS6d}}
\end{figure}

Taking $Pr_0^2 = 0.1$ and $a/r_0 = 1$ for the singly rotating Kerr-AdS$_6$ black hole, we
plot zero points of $\phi^{r_h}$ in the $r_h-\tau$ plane in Fig. \ref{6dKerrAdS}, and the
unit vector field $n$ with $\tau = 2.5r_0$ in Fig. \ref{KerrAdS6d}, respectively. For small
$\tau$, such as $\tau = \tau_1$, there are two intersection points for the singly rotating
Kerr-AdS$_6$ black hole. These two intersection points can coincide with each other when $\tau
= \tau_c$, and then vanish when $\tau > \tau_c$, therefore $\tau_c$ is an annihilation point
that can be found at $\tau_c = 2.81r_0$. Based upon the local property of the zero point, one
can easily obtain the the topological number $W = 0$ for the singly rotating Kerr-AdS$_6$
black hole, which is different from that of the six-dimensional singly rotating Kerr black
hole ($W = -1$) \cite{PRD107-024024}. This indicates that the cosmological constant is
important in determining the topological number for the six-dimensional rotating black hole.

\subsection{$d = 7$ case}\label{IVC}

Then, we consider $d = 7$ case with its generalized Helmholtz free energy being
\bea
\mathcal{F} &=& -\frac{3\pi^2r_h^2(r_h^2 +a^2)}{16\tau(15 -8\pi{}Pa^2)^2}
 \Big[20\pi{}r_h(15 -8\pi{}Pa^2) \nn \\
&&+\tau(8\pi{}Pr_h^2 +15)(8\pi{}Pa^2 -25) \Big] \, .
\eea
Therefore, one can straightforwardly obtain
\be
\tau = \frac{10\pi{}r_h(8\pi{}Pa^2 -15)(5r_h^2 +3a^2)}{(8\pi{}Pa^2 -25)
\big[6r_h^2(4\pi{}Pr_h^2 +5) +a^2(16\pi{}Pr_h^2 +15)\big]}
\ee
by solving the equation $\phi^{r_h} = 0$.

\begin{figure}[h]
\centering
\includegraphics[width=0.4\textwidth]{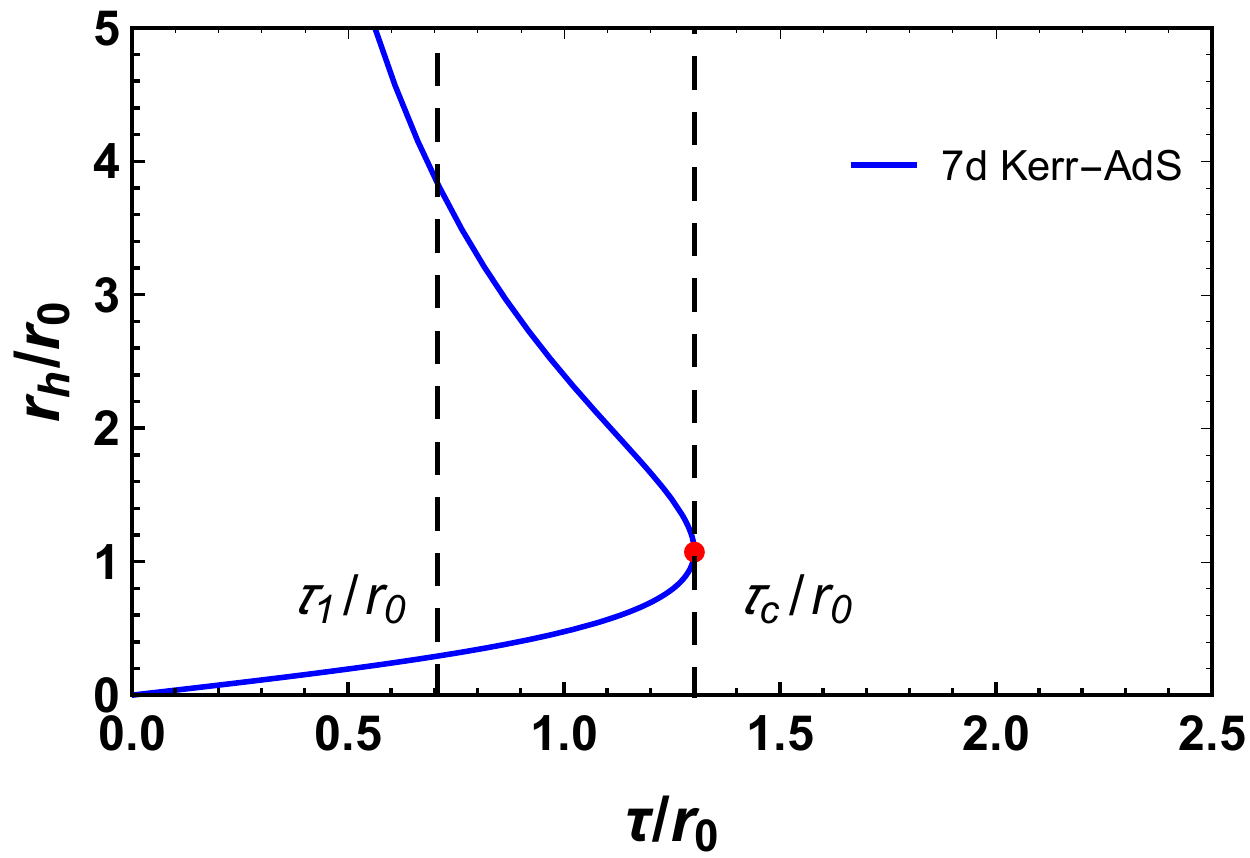}
\caption{Zero points of $\phi^{r_h}$ shown in the $r_h-\tau$ plane with
$Pr_0^2 = 0.3$ and $a/r_0 = 1$ for the singly rotating Kerr-AdS$_7$ black hole.
The red dot with $\tau_c$ denotes the annihilation point for the black hole.
There are two singly rotating Kerr-AdS$_7$ black holes for $\tau = \tau_1$.
Obviously, the topological number is $W = 0$. \label{7dKerrAdS}}
\end{figure}

\begin{figure}[h]
\centering
\includegraphics[width=0.35\textwidth]{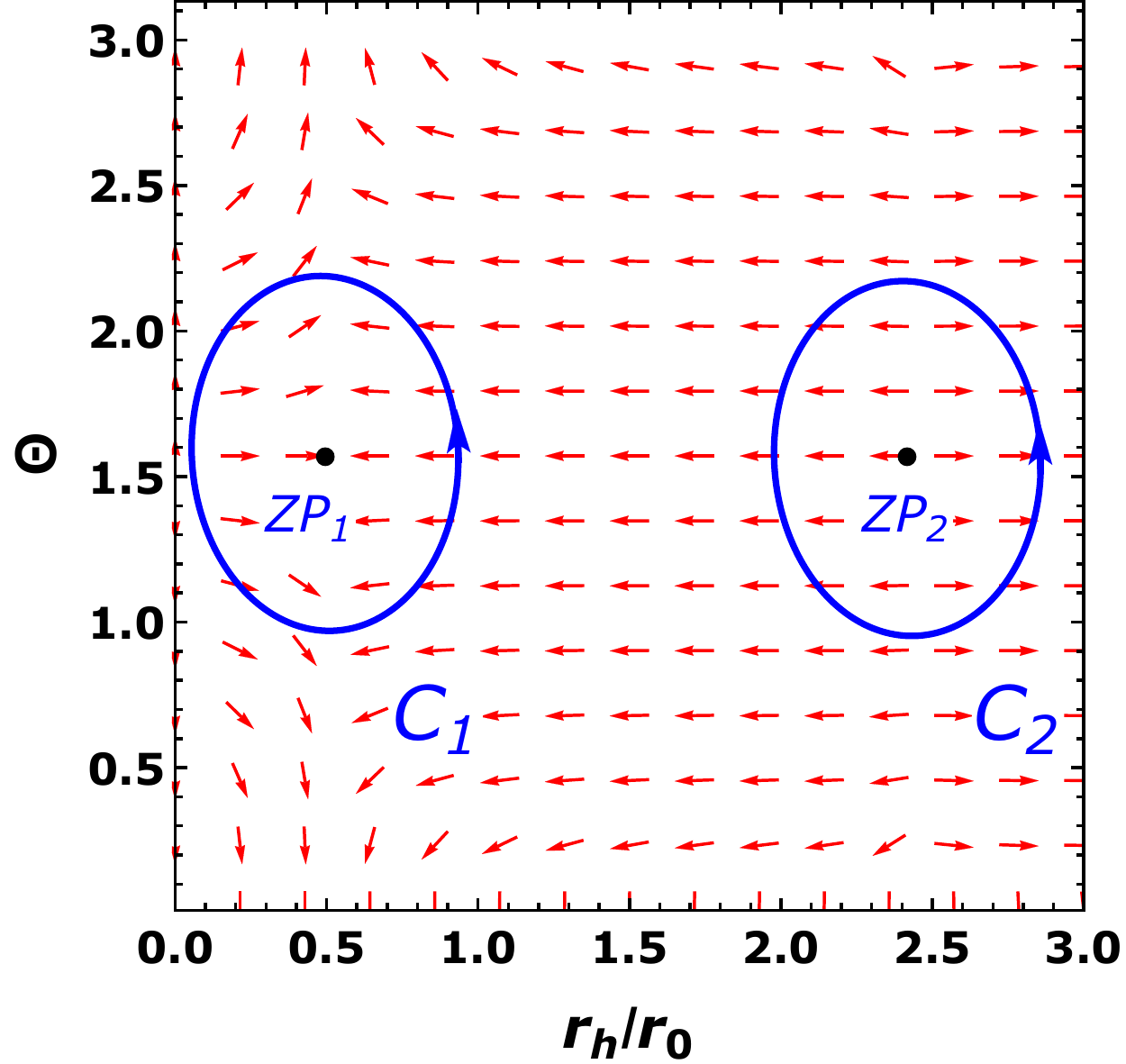}
\caption{The red arrows represent the unit vector field $n$ on a portion of the $r_h-\Theta$
plane with $Pr_0^2=0.3$, $a/r_0=1$, and $\tau=r_0$ for the singly rotating Kerr-AdS$_7$ black
hole. The zero points (ZPs) marked with black dots are at $(r_h/r_0, \Theta) = (0.48,\pi/2)$,
$(2.40,\pi/2)$ for ZP$_1$ and ZP$_2$, respectively. The blue contours $C_i$ are closed loops
surrounding the zero points. \label{KerrAdS7d}}
\end{figure}

Taking $Pr_0^2 = 0.3$ and $a/r_0 = 1$ for the singly rotating Kerr-AdS$_7$ black hole, we plot
the zero points of $\phi^{r_h}$ in the $r_h-\tau$ plane in Fig. \ref{7dKerrAdS}, and the unit
vector field $n$ with $\tau = r_0$ in Fig. \ref{KerrAdS7d}, respectively. Note that for the
values of $Pr_0^2 = 0.3$ and $a/r_0 = 1$, one annihilation point can be found at $\tau/r_0 =
\tau_c/r_0 = 1.30$. Based on the local property of the zero points, we get the topological number
$W = 0$ for the singly rotating Kerr-AdS$_7$ black hole, while that of the seven-dimensional
singly rotating Kerr black hole is: $W = -1$ \cite{PRD107-024024}. This demonstrates that the
cosmological constant is crucial to determine the topological number of the rotating black hole
in seven dimensions.

\subsection{$d = 8$ case}\label{IVD}

Let us turn to the $d = 8$ case. Similar to the procedure done in the previous three subsections,
one can get the generalized Helmholtz free energy as follows:
\bea
\mathcal{F} &=& -\frac{2\pi^2r_h^3(r_h^2 +a^2)}{15\tau(8\pi{}Pa^2 -21)^2}
 \Big[42\pi{}r_h(21 -8\pi{}Pa^2) \nn \\
&&+\tau(8\pi{}Pr_h^2 +21)(16\pi{}Pa^2 -63) \Big] \, .
\eea
As a result, by solving the equation $\phi^{r_h} = 0$, one can easily arrive at
\be
\tau = \frac{84\pi{}r_h(8\pi{}Pa^2 -21)(3r_h^2 +2a^2)}{(16\pi{}Pa^2 -63)
\big[56\pi{}Pr_h^4 +5(8\pi{}Pa^2 +21)r_h^2 +63a^2 \big]}
\ee
as the zero point of the vector field.

\begin{figure}[h]
\centering
\includegraphics[width=0.4\textwidth]{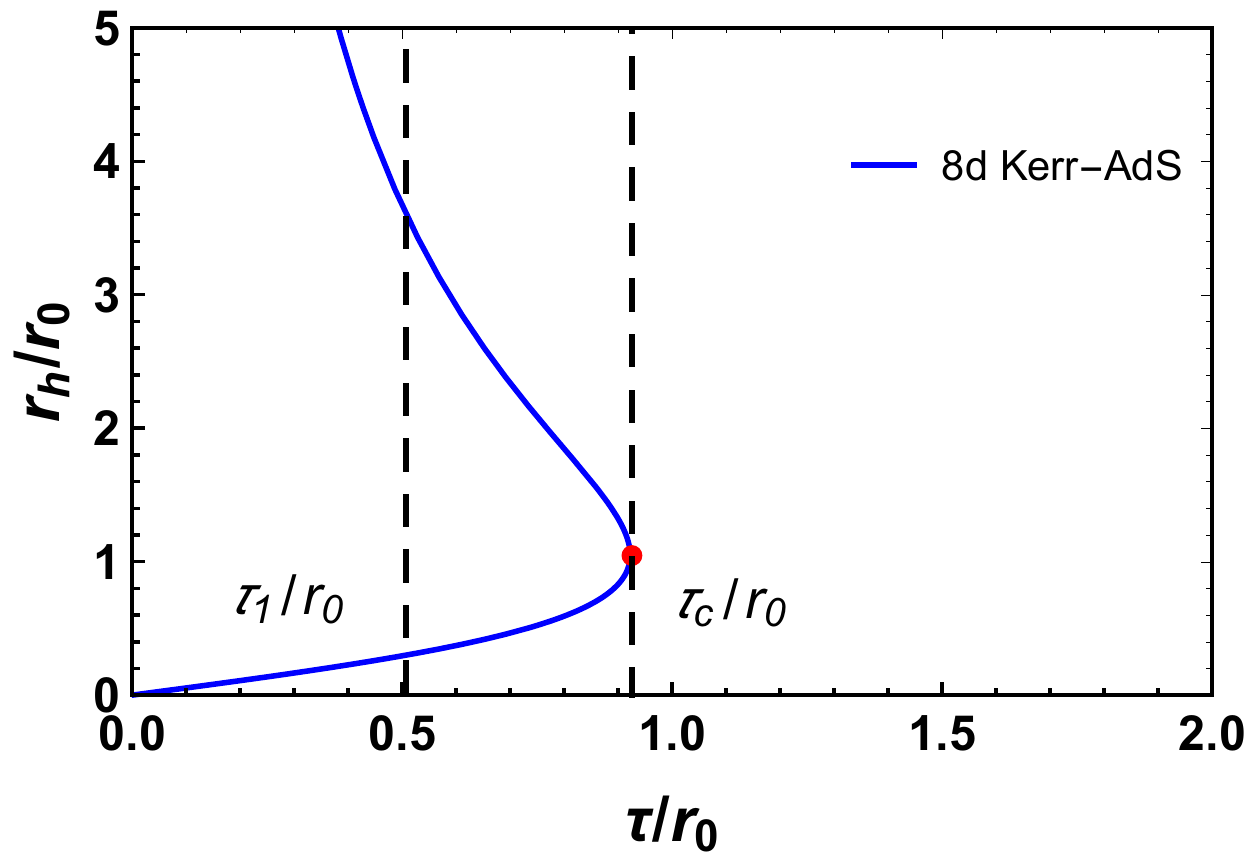}
\caption{Zero points of $\phi^{r_h}$ shown in the $r_h-\tau$ plane with $Pr_0^2=0.5$
and $a/r_0=1$ for the singly rotating eight-dimensional Kerr-AdS black hole. The red
dot with $\tau_c$ denotes the annihilation point for the black hole. There are two
singly rotating Kerr-AdS$_8$ black holes for $\tau = \tau_1$. It is easy to find that
the topological number is $W = 0$. \label{8dKerrAdS}}
\end{figure}

\begin{figure}[h]
\centering
\includegraphics[width=0.35\textwidth]{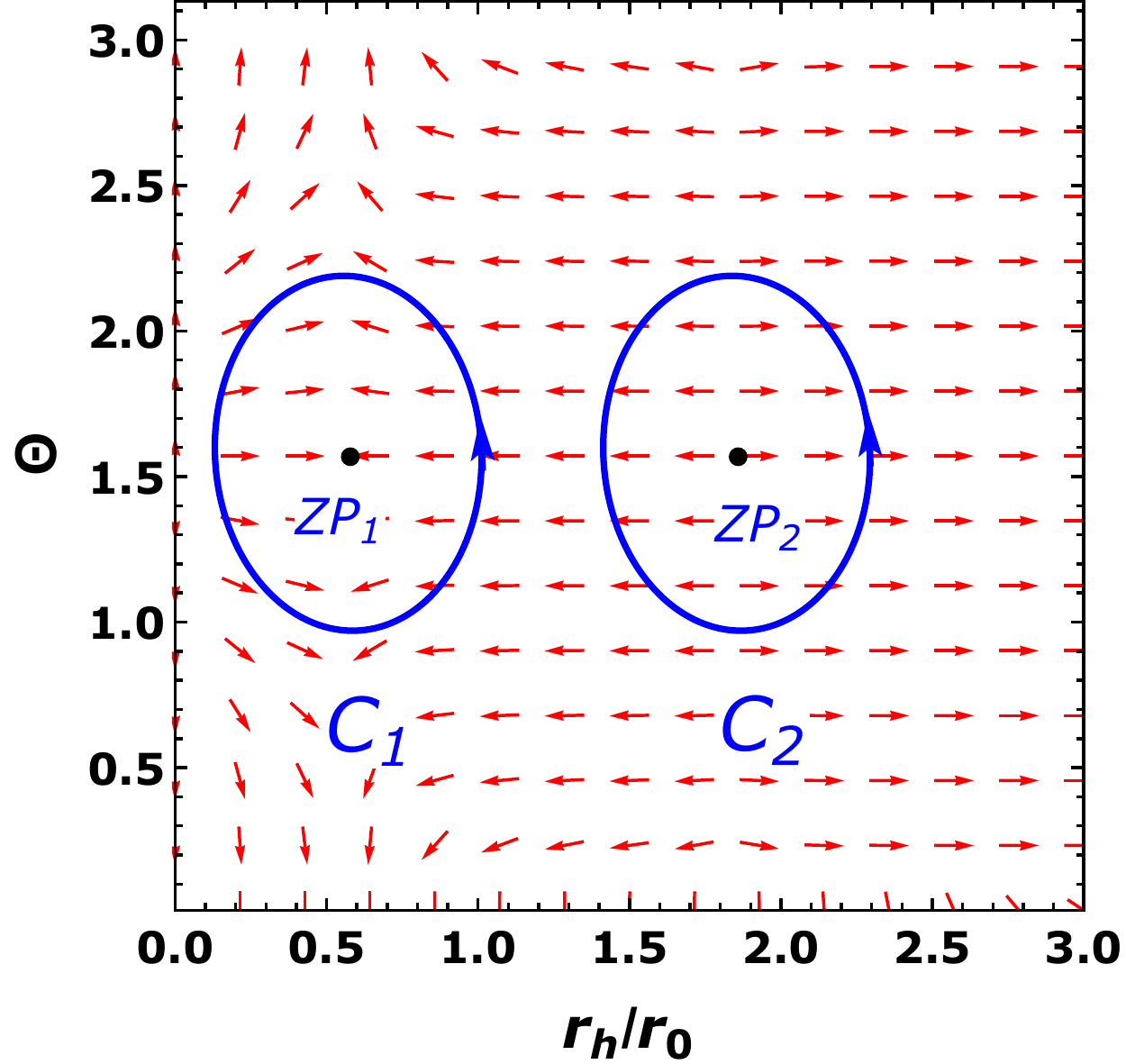}
\caption{The red arrows represent the unit vector field $n$ on a portion of the $r_h-\Theta$
plane with $Pr_0^2=0.5$, $a/r_0=1$, and $\tau/r_0=0.8$ for the singly rotating Kerr-AdS$_8$
black hole. The zero points (ZPs) marked with black dots are at $(r_h/r_0, \Theta) = (0.59,
\pi/2)$, $(1.85,\pi/2)$ for ZP$_1$ and ZP$_2$, respectively. The blue contours $C_i$ are
closed loops surrounding the zero points. \label{KerrAdS8d}}
\end{figure}

In Figs. \ref{8dKerrAdS} and \ref{KerrAdS8d}, taking $Pr_0^2 = 0.5$ and $a/r_0 = 1$ for
the singly rotating Kerr-AdS$_8$ black hole, we plot the zero points of $\phi^{r_h}$ in
the $r_h-\tau$ plane and the unit vector field $n$ with $\tau/r_0 = 0.8$, respectively.
Note that for the values of $Pr_0^2 = 0.5$ and $a/r_0 = 1$, one annihilation point can
be found at $\tau/r_0 = \tau_c/r_0 = 0.92$. Based on the local property of the zero points,
it is easy to find that the topological number $W = 0$ for the singly rotating Kerr-AdS$_8$
black hole. Combined with the fact that the eight-dimensional singly rotating Kerr black hole
has a topological number: $W = -1$ \cite{PRD107-024024}, it is evident that the cosmological
constant is important in  determining the topological number for the rotating black hole in
eight dimensions.

\subsection{$d = 9$ case}\label{IVE}

Finally, we investigate the topological number for the nine-dimensional singly rotating
Kerr-AdS black hole whose generalized Helmholtz free energy is
\bea
\mathcal{F} &=& -\frac{\pi^3r_h^4(r_h^2 +a^2)}{48\tau(2\pi{}Pa^2 -7)^2}
\Big[28\pi{}r_h(7 -2\pi{}Pa^2) \nn \\
&&+\tau(10\pi{}Pa^2 -49)(2\pi{}Pr_h^2 +7) \Big] \, .
\eea
Thus, the zero point of the vector constructed in the topological approach can be written as
\be
\tau = \frac{14\pi{}r_h(2\pi{}Pa^2 -7)(7r_h^2 +5a^2)}{(10\pi{}Pa^2 -49)[8\pi{}Pr_h^4
 +3(2\pi{}Pa^2 +7)r_h^2 +14a^2]} \, .
\ee

\begin{figure}[h]
\centering
\includegraphics[width=0.4\textwidth]{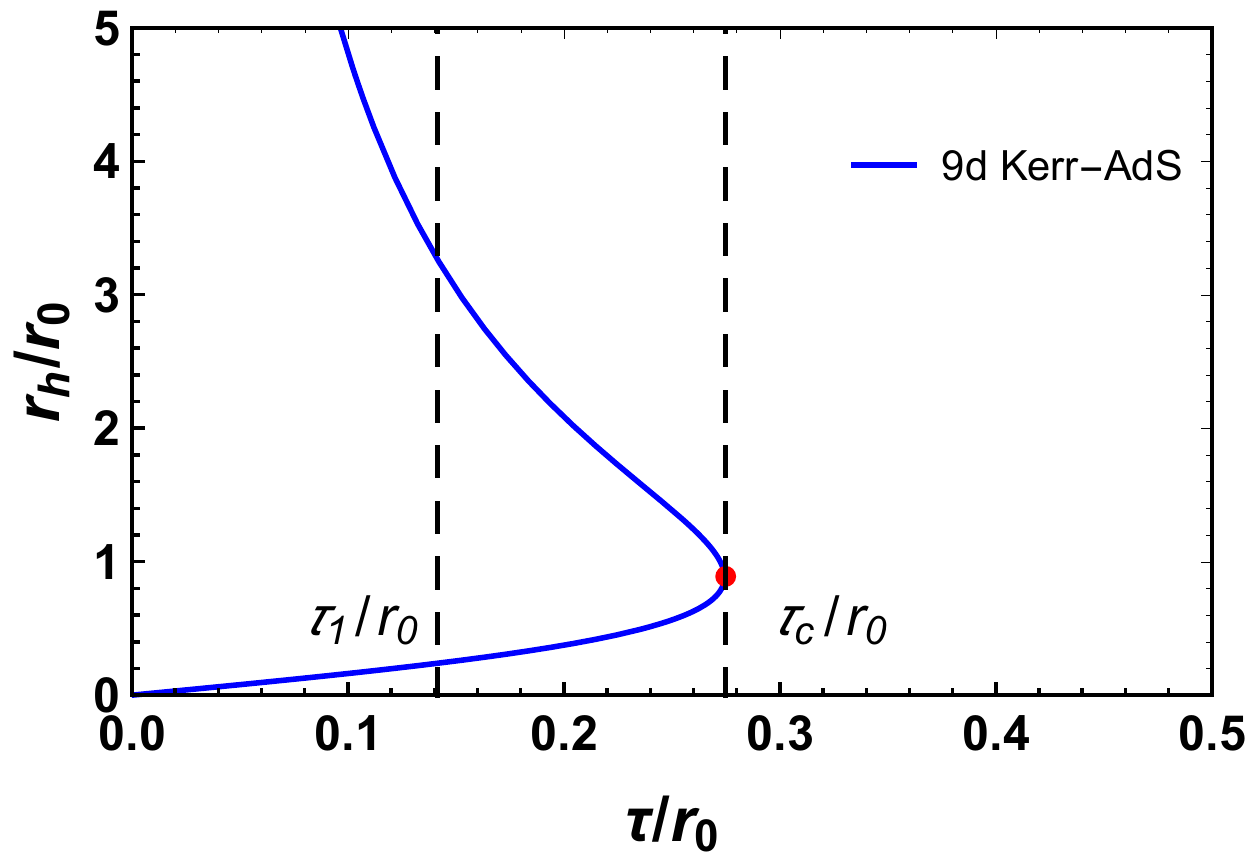}
\caption{Zero points of $\phi^{r_h}$ shown in the $r_h-\tau$ plane with $Pr_0^2=1$
and $a/r_0=1$ for the singly rotating Kerr-AdS$_9$ black hole. The red dot with
$\tau_c$ represents the annihilation point for the black hole. There are two singly
rotating Kerr-AdS$_9$ black holes when $\tau = \tau_1$. Furthermore, the topological
number is $W = 0$. \label{9dKerrAdS}}
\end{figure}

\begin{figure}[h]
\centering
\includegraphics[width=0.35\textwidth]{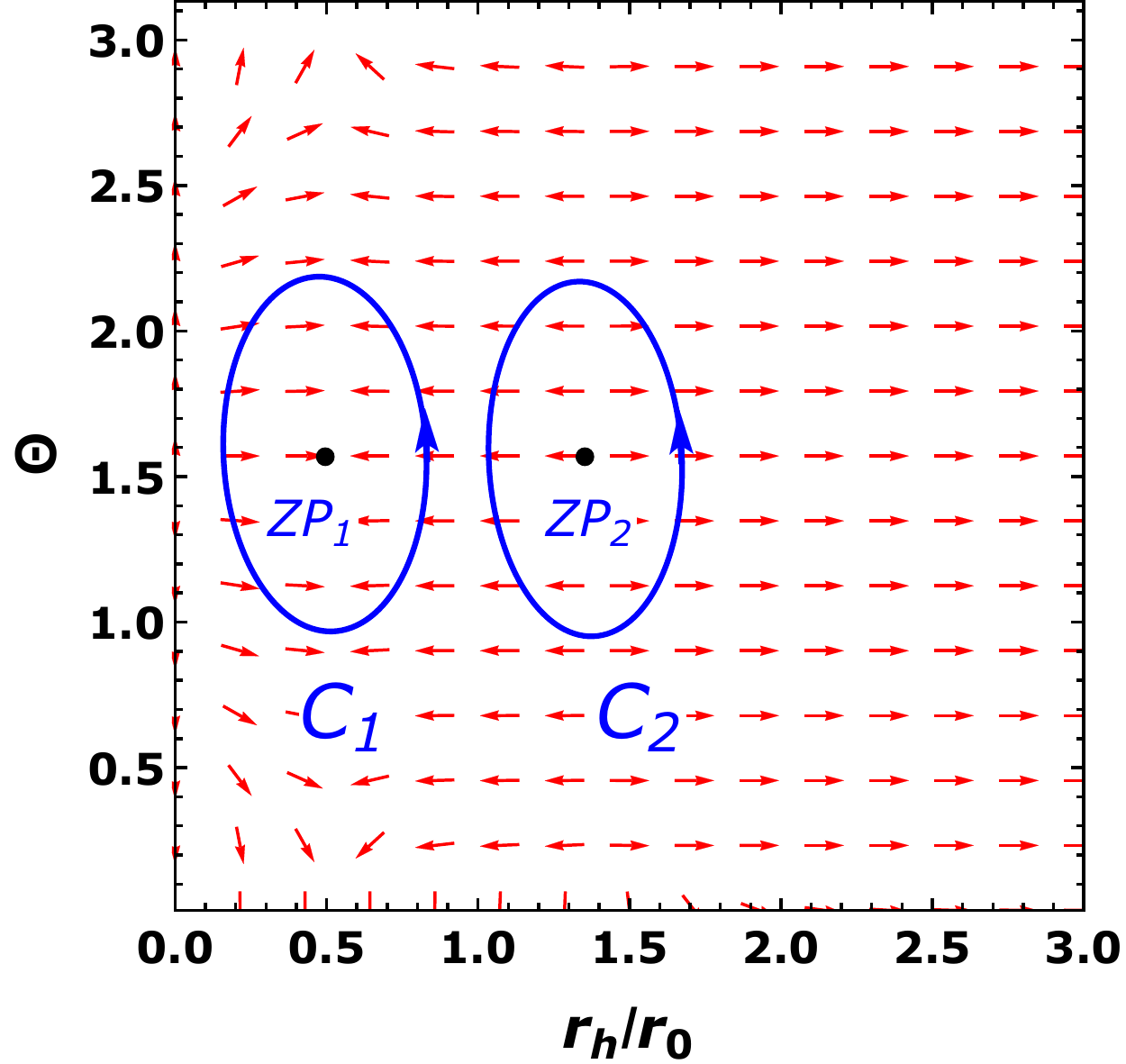}
\caption{The red arrows represent the unit vector field $n$ on a portion of the $r_h-\Theta$
plane with $Pr_0^2=1$, $a/r_0=1$, and $\tau/r_0=0.25$ for the singly rotating Kerr-AdS$_9$
black hole. The zero points (ZPs) marked with black dots are at $(r_h/r_0, \Theta) = (0.56,
\pi/2)$, $(1.39,\pi/2)$ for ZP$_1$ and ZP$_2$, respectively. The blue contours $C_i$ are
closed loops surrounding the zero points. \label{KerrAdS9d}}
\end{figure}

In Figs. \ref{9dKerrAdS} and \ref{KerrAdS9d}, taking $Pr_0^2 = 1$ and $a/r_0 = 1$ for the
nine-dimensional singly rotating Kerr-AdS black hole, we plot the zero points of $\phi^{r_h}$
in the $r_h-\tau$ plane and the unit vector field $n$ with $\tau/r_0 = 0.25$, respectively.
Note that for the values of $Pr_0^2 = 1$ and $a/r_0 = 1$, one annihilation point can be
found at $\tau/r_0 = \tau_c/r_0 = 0.27$. Based on the local property of the zero points,
the topological number is easily determined as $W = 0$, which is different from the topological
number of the nine-dimensional singly rotating Kerr black hole ($W = -1$) \cite{PRD107-024024}.
Therefore, this fact indicates that the cosmological constant plays a significant role in
determining the topological number for the nine-dimensional rotating black hole.

\subsection{Summary: The impact of dimension of the spacetime}

Summarizing our results from Subsects. \ref{IVA}-\ref{IVE}, we can find that the topological
number of the five-dimensional singly rotating Kerr-AdS$_5$ black hole is $W = 0$, while the
six- to nine-dimensional singly rotating Kerr-AdS black holes have both $W = -1$. Thus it
indicates the dimension of spacetime has an important effect on the topological number for
rotating AdS black holes.

\section{Three-dimensional rotating BTZ black hole}\label{V}

Because the BTZ black hole \cite{PRL69-1849,PRD48-1506} is a first nontrivial exact solution
to the three-dimensional gravity theory, it is important to study the topological number
of the rotating BTZ black hole. Therefore, in this section, we turn our attention to the
three-dimensional BTZ black hole solution, whose metric is given by \cite{PRL69-1849,
PRD48-1506,CQG12-2853,PRD59-064005}
\be\label{rotBTZ}
ds^2 = -f(r)dt^2 +\frac{dr^2}{f(r)} +r^2\Big(d\varphi -\frac{J}{2r^2}dt \Big)^2 \, ,
\ee
where
\be
f(r) = -2m +\frac{r^2}{l^2} +\frac{J^2}{4r^2} \, , \nn
\ee
in which $m$ is the mass parameter, $l$ is the AdS radius, $J$ is the angular momentum that
must satisfy $|J|\le ml$.

The mass and entropy associated with the above solution (\ref{rotBTZ}) are given by
\cite{PRD92-124069}
\be
M = \frac{m}{4} = \frac{r_h^2}{8l^2} +\frac{J^2}{32r_h^2} \, , \qquad
S = \frac{1}{2}\pi{}r_h \, ,
\ee
where $r_h$ is the location of the event horizon. Utilizing the definition of the generalized
Helmholtz free energy (\ref{FE}) and substituting $l^2 = 1/(8\pi{}P)$, one can obtain
\be
\mathcal{F} = \pi{}Pr_h^2 +\frac{J^2}{32r_h^2} -\frac{\pi{}r_h}{2\tau} \, .
\ee
Thus, the components of the vector $\phi$ are
\be
\phi^{r_h} = 2\pi{}Pr_h -\frac{J^2}{16r_h^3} -\frac{\pi}{2\tau} \, , \quad
\phi^{\Theta} = -\cot\Theta\csc\Theta \, .
\ee
By solving the equation $\phi^{r_h} = 0$, one can obtain
\be
\tau = \frac{8\pi{}r_h^3}{32\pi{}Pr_h^4 -J^2}
\ee
as the zero point of the vector field $\phi$.

\begin{figure}[h]
\centering
\includegraphics[width=0.4\textwidth]{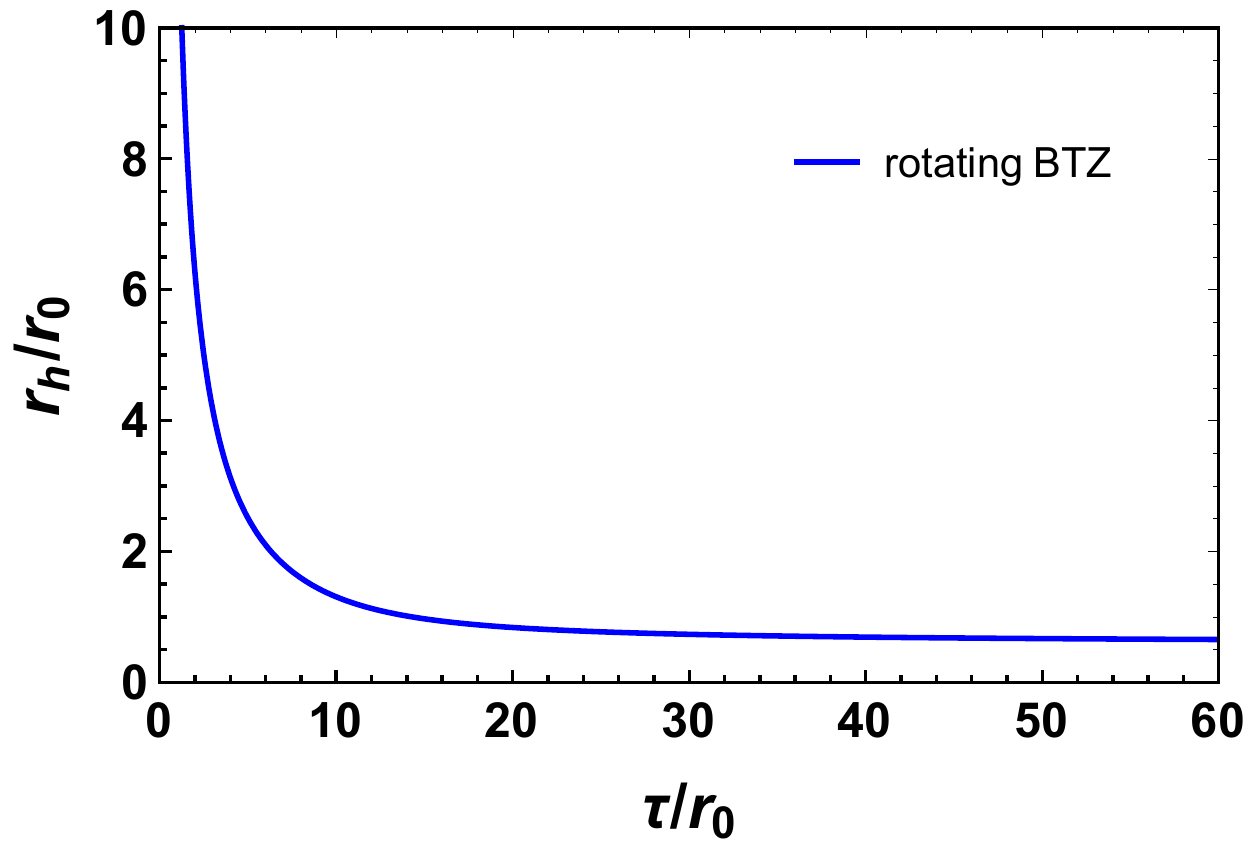}
\caption{Zero point of the vector $\phi^{r_h}$ shown on the $r_h-\tau$ plane with $Pr_0^2 = 0.02$
and $J/r_0 = 0.5$ for the rotating BTZ black hole. There is only one thermodynamically stable
rotating BTZ black hole for any value of $\tau$. \label{3drotBTZ}}
\end{figure}

\begin{figure}[h]
\centering
\includegraphics[width=0.35\textwidth]{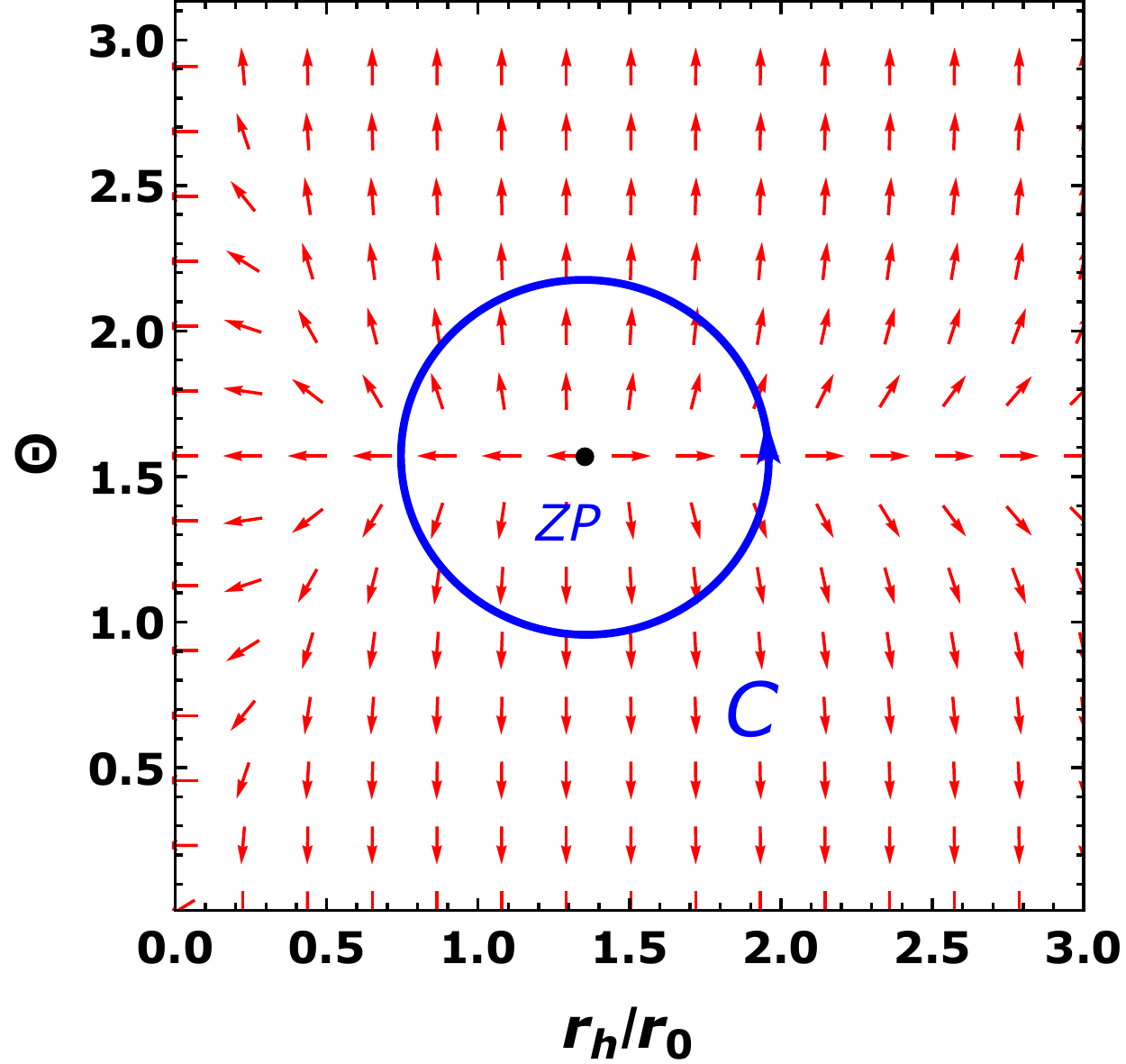}
\caption{The red arrows represent the unit vector field $n$ on a portion of the $r_h-\Theta$
plane with $Pr_0^2=0.02$, $J/r_0=0.5$, and $\tau/r_0=10$ for the rotating BTZ black hole. The
zero point (ZP) marked with black dot is at $(r_h/r_0, \Theta) = (1.31,\pi/2)$. The blue contour
$C$ is closed loop surrounding the zero point. \label{rotBTZ3d}}
\end{figure}

For the three-dimensional rotating BTZ black hole, we take $Pr_0^2=0.02$ and $J/r_0=0.5$, and
then plot the zero points of the component $\phi^{r_h}$ in Fig. \ref{3drotBTZ}, and the unit
vector field $n$ with $\tau/r_0=10$ in Fig. \ref{rotBTZ3d}, respectively. Obviously, there is
only one thermodynamically stable rotating BTZ black hole for any value of $\tau$, which is
also consistent with the conclusion given in Ref. \cite{PRD104-124003} via the Joule-Thomson
expansion. Based on the local property of the zero points, the topological number $W = 1$ can
be found for the three-dimensional rotating BTZ black hole. In the Appendix \ref{App}, we will
also investigate the topological number of the three dimensional charged BTZ black hole, and
find that its value is: $W = 0$.

\section{Kerr-Newman-AdS$_4$ black hole}\label{VI}

Finally, we would like to investigate the topological number of the four-dimensional
Kerr-Newman-AdS black hole \cite{CMP10-280}, whose metric and Abelian gauge potential
are \cite{CQG17-399,PRD59-064005}
\bea\label{KNAdS}
ds^2 &=& -\frac{\Delta_r}{\Sigma}\Big(\frac{\Delta_\theta}{\Xi}dt
 -\frac{a}{\Xi}\sin^2\theta{}d\varphi \Big)^2
+\frac{\Sigma}{\Delta_r}dr^2 +\frac{\Sigma}{\Delta_\theta}d\theta^2 \nn \\
&&+\frac{\Delta_\theta\sin^2\theta}{\Sigma}\Big[\frac{a(r^2 +l^2)}{l^2\Xi}dt
 -\frac{r^2 +a^2}{\Xi}d\varphi \Big]^2 \, , \\
A &=& \frac{qr}{\Sigma}\Big(\frac{\Delta_\theta}{\Xi}dt
 -\frac{a\sin^2\theta}{\Xi} d\varphi \Big) \, ,
\eea
where
\bea
&&\Delta_r = (r^2 +a^2)\Big(1 +\frac{r^2}{l^2}\Big) -2mr +q^2 \, , \quad
\Xi = 1 -\frac{a^2}{l^2} \, , \nn \\
&&\Delta_\theta = 1 -\frac{a^2}{l^2}\cos^2\theta \, , \quad
\Sigma = r^2 +a^2\cos^2\theta \, , \nn
\eea
in which $a$, $m$ and $q$ are the rotation, the mass and electric charge parameters, respectively,
and $l$ is the AdS radius. The horizon radius $r_h$ is determined by equation: $\Delta_r = 0$.

The mass and entropy associated with the above metric (\ref{KNAdS}) can be calculated via the
standard method and their results are
\be
M = \frac{m}{\Xi^2} \, , \qquad S = \frac{\pi(r_h^2 +a^2)}{\Xi} \, .
\ee
Then, one can straightforwardly obtain the generalized Helmholtz free energy of this black hole
as
\bea
\mathcal{F} &=& \frac{24\pi{}Pr_h^2(r_h^2 +a^2) +a^2[16\pi{}PQ^2(4\pi{}Pa^2 -3)
 +9]}{2r_h(8\pi{}Pa^2 -3)^2} \nn \\
&&+\frac{9(r_h^2 +Q^2)}{2r_h(8\pi{}Pa^2 -3)^2}
 +\frac{6\pi{}(r_h^2 +a^2)}{2\tau(8\pi{}Pa^2 -3)} \, , \quad
\eea
with $Q = q/\Xi$ being the electric charge of the black hole. Therefore, the zero point of the vector
can be easily given as
\be
\tau = \frac{12\pi{}r_h^3(8\pi{}Pa^2 -3)}{X -24\pi{}P(3r_h^2 +a^2)r_h^2} \, ,
\ee
where $X = a^2[16\pi{}PQ^2(4\pi{}Pa^2 -3) +9] +9(Q^2 -r_h^2)$.

\begin{figure}[h]
\centering
\includegraphics[width=0.4\textwidth]{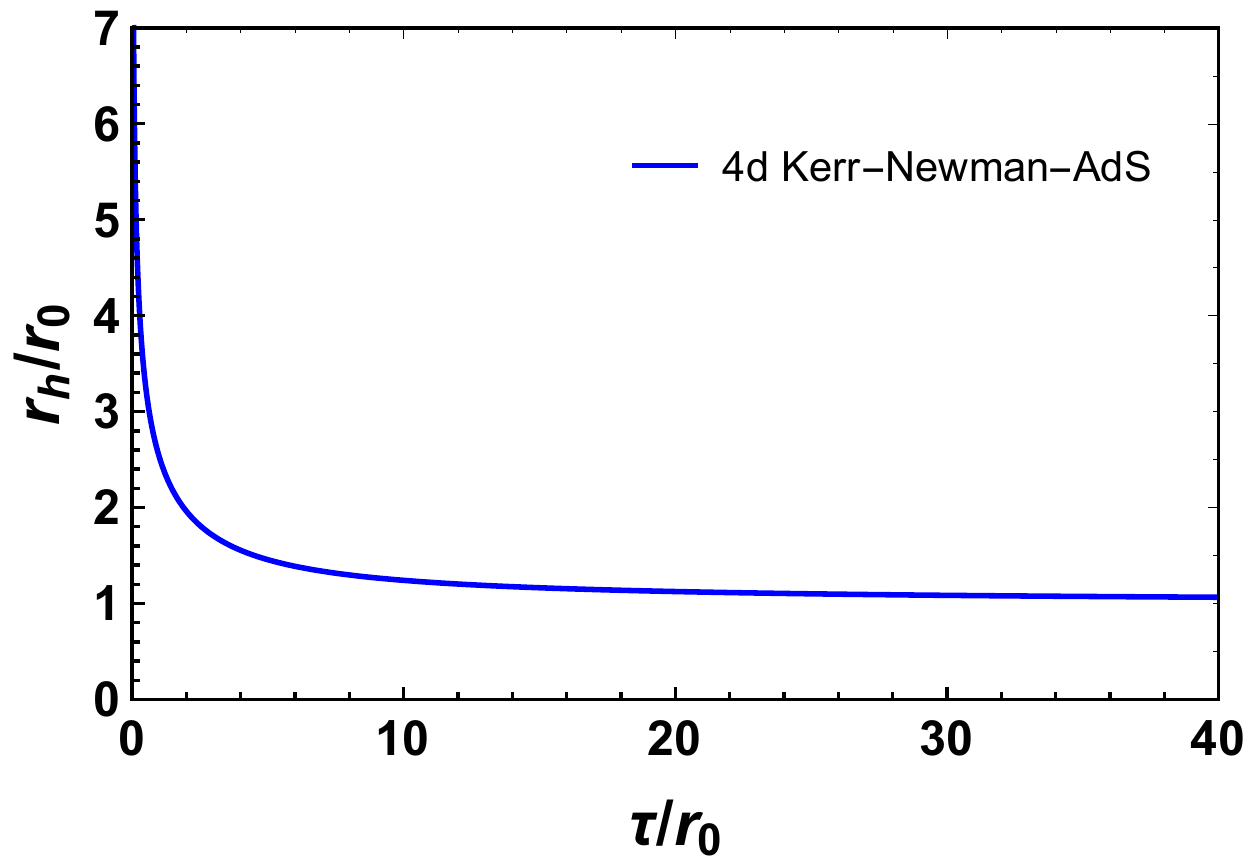}
\caption{Zero point of the vector $\phi^{r_h}$ shown on the $r_h-\tau$ plane with
$Pr_0^2 = 0.02$, $a/r_0 = 1$, and $Q/r_0 = 1$ for the Kerr-Newman-AdS$_4$ black hole.
There is only one stable Kerr-Newman-AdS$_4$ black hole for any value of $\tau$. The
topological number of this black hole is $W = 1$.
\label{4dKNAdS}}
\end{figure}

\begin{figure}[h]
\centering
\includegraphics[width=0.35\textwidth]{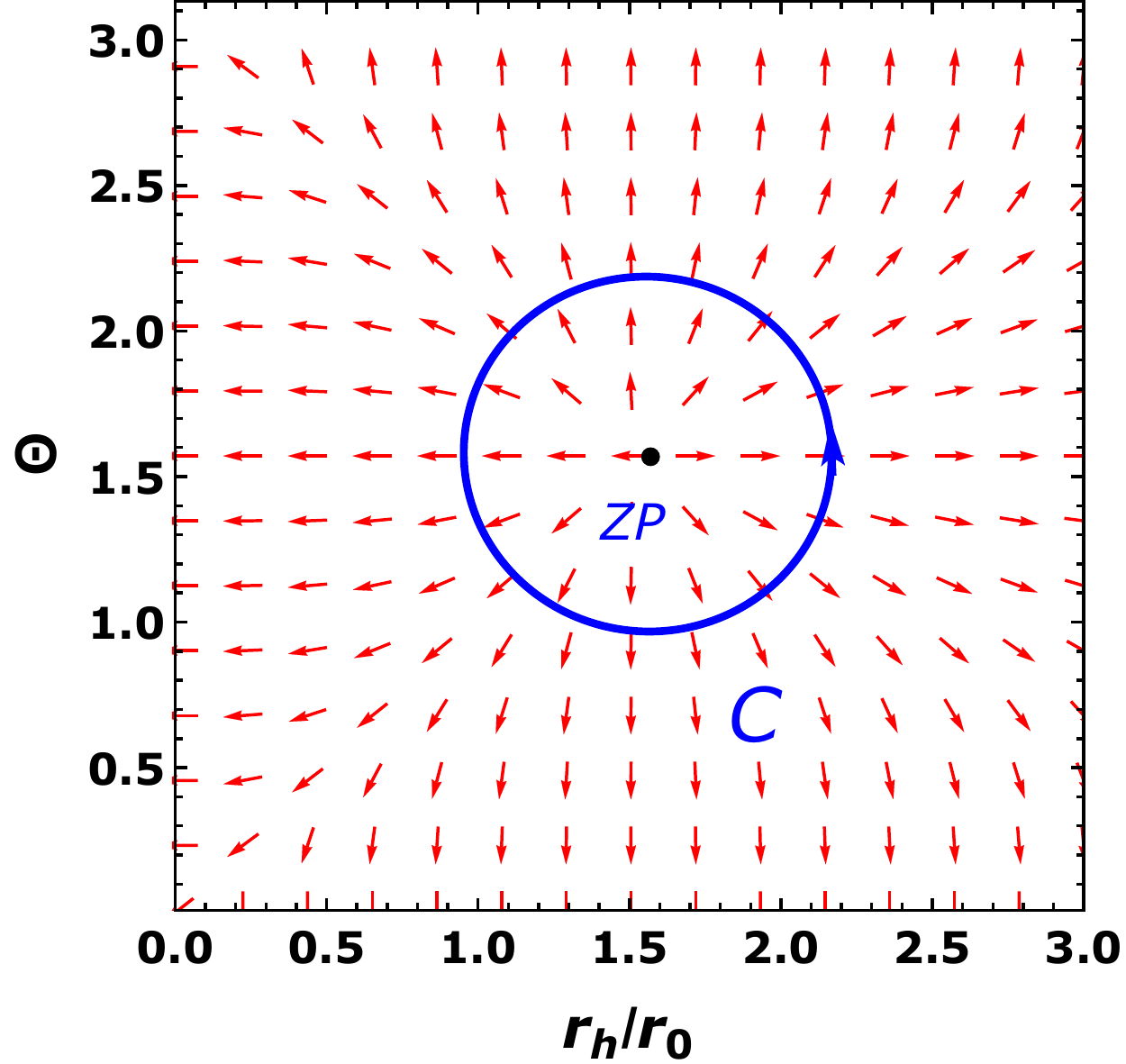}
\caption{The red arrows represent the unit vector field $n$ on a portion of the $r_h-\Theta$
plane with $Pr_0^2=0.02$, $a/r_0=1$, $Q/r_0=1$ and $\tau/r_0=10$ for the Kerr-Newman-AdS$_4$
black hole. The zero point (ZP) marked with black dot is at $(r_h/r_0,\Theta) = (1.52,\pi/2)$.
The blue contour $C$ is closed loop surrounding the zero point. \label{KNAdS4d}}
\end{figure}

For the four-dimensional Kerr-Newman-AdS black hole, we take $Pr_0^2=0.02$, $a/r_0=1$, $Q/r_0=1$,
and plot the zero points of the component $\phi^{r_h}$ in Fig. \ref{4dKNAdS}, and the unit vector
field $n$ with $\tau/r_0=10$ in Fig. \ref{KNAdS4d}, respectively. Obviously, there is only one
thermodynamically stable Kerr-Newman-AdS$_4$ black hole for any value of $\tau$. Based upon the
local property of the zero points, one can get the topological number $W = 1$ for the four-dimensional
Kerr-Newman-AdS black hole, which is identical to that of the four-dimensional Kerr-AdS black hole.
This fact indicates that the electric charge parameter has no effect on the topological number of
rotating AdS black holes. Compared with the four-dimensional Kerr-Newman black hole which has a
topological number of zero, it can be inferred that the cosmological constant plays an crucial
role in determining the topological number for the rotating charged black hole.

\section{Conclusions}\label{VII}

In this paper, we have extended our previous work \cite{PRD107-024024} to the more general rotating
AdS black hole cases and investigated the topological numbers of the singly rotating Kerr-AdS black
holes in arbitrary dimensions and the four-dimensional Kerr-Newman-AdS black hole as well as the
three-dimensional rotating or charged BTZ black hole. Table \ref{TableI} summarizes some interesting
results found in the present work. We find that the $d\ge 6$ singly rotating Kerr-AdS black holes,
the Schwarzschild-AdS black hole, and the charged BTZ black hole belong to the same kind of topological
classes because of their topological numbers being $W = 0$, while the Kerr-Newman-AdS black hole,
the $d = 4, 5$ singly rotating Kerr-AdS black holes, and the rotating BTZ black hole belong to
another kind of topological class due to their topological number being $W = 1$.

\begin{table}[h]
\caption{The topological number $W$, numbers of generation and annihilation points for various
AdS black holes.}
\resizebox{0.48\textwidth}{!}{
\begin{tabular}{c|c|c|c}
\hline\hline
BH solution & $W$ & Generation point &Annihilation point\\ \hline
Schwarzschild-AdS$_4$ BH \cite{PRD106-064059} & 0 & 0 & 1\\
$d\ge 6$ singly rotating Kerr-AdS BH & 0 & 0 & 1\\
Charged BTZ BH & 0& 0 & 1\\ \hline
$d=5$ singly rotating Kerr-AdS BH & 1 & 1 or 0 & 1 or 0\\
RN-AdS$_4$ BH \cite{PRL129-191101} & 1 & 1 or 0 & 1 or 0\\
Kerr-AdS$_4$ BH & 1 & 1 or 0 & 1 or 0\\
Kerr-Newman-AdS$_4$ BH & 1 & 0 & 0\\
Rotating BTZ BH & 1 & 0 & 0\\
\hline\hline
\end{tabular}}
\label{TableI}
\end{table}

As a new consequence, we have discovered that the topological number of the rotating black holes
is significantly influenced by the cosmological constant. Furthermore, combining our results with
those in Refs. \cite{PRL129-191101,PRD106-064059,PRD107-024024}, we have tabulated Table \ref{TableII},
from which we have also observed a new interesting phenomenon: the difference between the topological
number of the AdS black hole and that of its corresponding asymptotically flat black hole is always
unity. We conjecture that this might also be true for other kinds of black holes. However, this
needs to be tested by further investigating the topological numbers of many other black holes and
their AdS counterparts.

\begin{table}[h]
\caption{The topological number $W$, numbers of generation and annihilation points for various
black holes and their AdS extensions.}
\resizebox{0.48\textwidth}{!}{
\begin{tabular}{c|c|c|c}
\hline\hline
BH solution & $W$ & Generation point &Annihilation point\\ \hline
Schwarzschild BH \cite{PRL129-191101} & -1 & 0 & 0\\
Schwarzschild-AdS$_4$ BH \cite{PRD106-064059} & 0 & 0 & 1\\ \hline
RN BH \cite{PRL129-191101} & 0 & 1 & 0 \\
RN-AdS$_4$ BH \cite{PRL129-191101} & 1 & 1 or 0 & 1 or 0\\ \hline
Kerr BH \cite{PRD107-024024} & 0 & 1 & 0\\
Kerr-AdS$_4$ BH & 1 & 1 or 0 & 1 or 0\\ \hline
Kerr-Newman BH \cite{PRD107-024024} & 0 & 1 & 0\\
Kerr-Newman-AdS$_4$ BH & 1 & 0 & 0\\ \hline
$d=5$ singly rotating Kerr BH \cite{PRD107-024024} & 0 & 1 & 0\\
$d=5$ singly rotating Kerr-AdS BH & 1 & 1 or 0 & 1 or 0\\ \hline
$d\ge 6$ singly rotating Kerr BH \cite{PRD107-024024} & -1 & 0 & 0\\
$d\ge 6$ singly rotating Kerr-AdS BH & 0 & 0 & 1\\
\hline\hline
\end{tabular}}
\label{TableII}
\end{table}

As far as the impact of the electric charge parameter on the topological number of the
four-dimensional black holes is concerned, one can infer from Table \ref{TableI} that the
electric charge parameter can change the topological number of the static AdS$_4$ black holes,
while from Tables \ref{TableI} and \ref{TableII} that it has no impact on the topological number of
the four-dimensional rotating AdS black holes because the four-dimensional Kerr-AdS and
Kerr-Newman-AdS black holes have the same topological number.

Finally, it should be mentioned that all rotating AdS black holes studied in the present paper
are under-rotating, namely, their rotation angular velocities are less than the speed of light
(i.e., $a < l$). Therefore, a most related issue is to investigate their ultraspinning and
over-rotating cases, and to test our guess by checking the topological numbers of the ultraspinning
AdS black holes (i.e., $a = l$) \cite{PRD89-084007, PRL115-031101,JHEP0114127,PRD103-104020,
PRD101-024057,PRD102-044007,PRD103-044014,JHEP1121031} and the over-rotating Kerr-AdS black
holes (i.e., $a > l$) \cite{CQG38-095001,PRD103-024053}.

\acknowledgments

We thank Prof. Yen Chin Ong for useful advices. We are also greatly indebted to the anonymous
referee for his/her constructive comments to improve the presentation of this work. This work
is supported by the National Natural Science Foundation of China (NSFC) under Grant No. 12205243,
No. 11675130, by Sichuan Science and Technology Program under Grant No. 2023NSFSC1347,
and by the Doctoral Research Initiation Project of China West Normal University under Grant No.
21E028.

\appendix

\section{Three-dimensional charged BTZ black hole}\label{App}

In this appendix, we will investigate the topological number of the three-dimensional charged
BTZ black hole, whose metric reads \cite{PRD61-104013}
\bea
ds^2 &=& -f(r)dt^2 +\frac{dr^2}{f(r)} +r^2d\varphi^2 \, , \\
A &=& -q\ln\Big(\frac{r}{l}\Big)dt \, ,
\eea
where
\be
f(r) = -2m -\frac{q^2}{2}\ln\Big(\frac{r}{l}\Big) +\frac{r^2}{l^2} \, ,
\ee
in which $m$ and $q$ are the mass parameter and electric charge, respectively, and $l$ is
the AdS radius. The event horizon is determined by: $f(r_h) = 0$.

For the three-dimensional charged BTZ black hole, the mass and the entropy
are \cite{PRD92-124069}
\be
M = \frac{m}{4} = \frac{r_h^2}{8l^2} -\frac{q^2}{16}\ln\Big(\frac{r_h}{l}\Big) \, , \qquad
S = \frac{1}{2}\pi{}r_h \, .
\ee

\begin{figure}[h]
\centering
\includegraphics[width=0.4\textwidth]{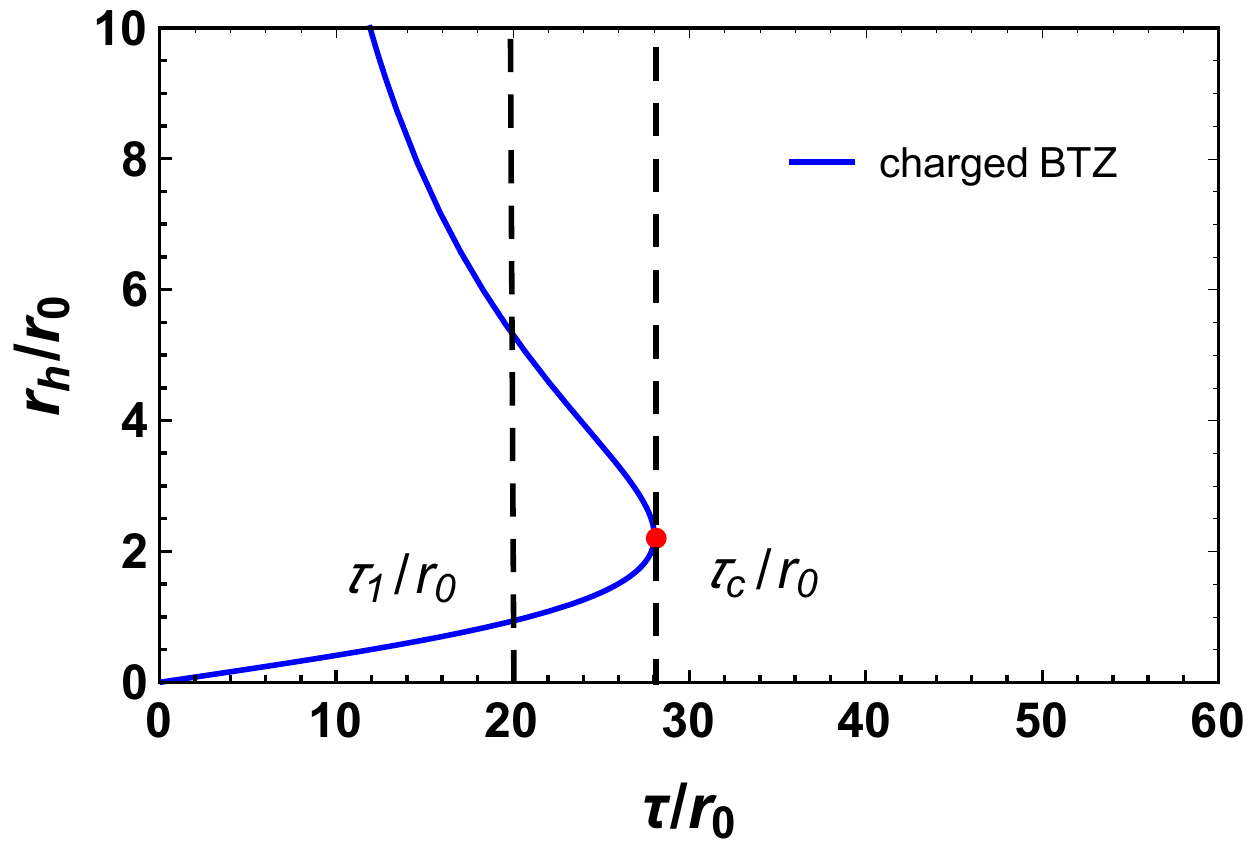}
\caption{Zero points of $\phi^{r_h}$ shown in the $r_h-\tau$ plane with $P=0.002$ and
$q/r_0=1$ for the charged BTZ black hole. The red dot with $\tau_c$ represents the
annihilation point for the black hole. There are two charged BTZ black holes when
$\tau = \tau_1$. Furthermore, its topological number is: $W = 0$. \label{3dchargedBTZ}}
\end{figure}

Substituting $l^2 = 1/(8\pi{}P)$  into the definition of the generalized Helmholtz free
energy (\ref{FE}), one can arrive at
\be
\mathcal{F} = \pi{}Pr_h^2 +\frac{q^2}{16}\ln(2r_h\sqrt{2\pi{}P}) -\frac{\pi{}r_h}{2\tau} \, ,
\ee
Thus, the components of the vector $\phi$ are
\be
\phi^{r_h} = 2\pi{}Pr_h -\frac{q^2}{16r_h^3} -\frac{\pi}{2\tau} \, , \quad
\phi^{\Theta} = -\cot\Theta\csc\Theta \, .
\ee
By solving the equation $\phi^{r_h} = 0$, one can obtain
\be\label{tauA}
\tau = \frac{8\pi{}r_h}{32\pi{}Pr_h^2 +q^2}
\ee
as the zero point of the vector field $\phi$.

\begin{figure}[h]
\centering
\includegraphics[width=0.35\textwidth]{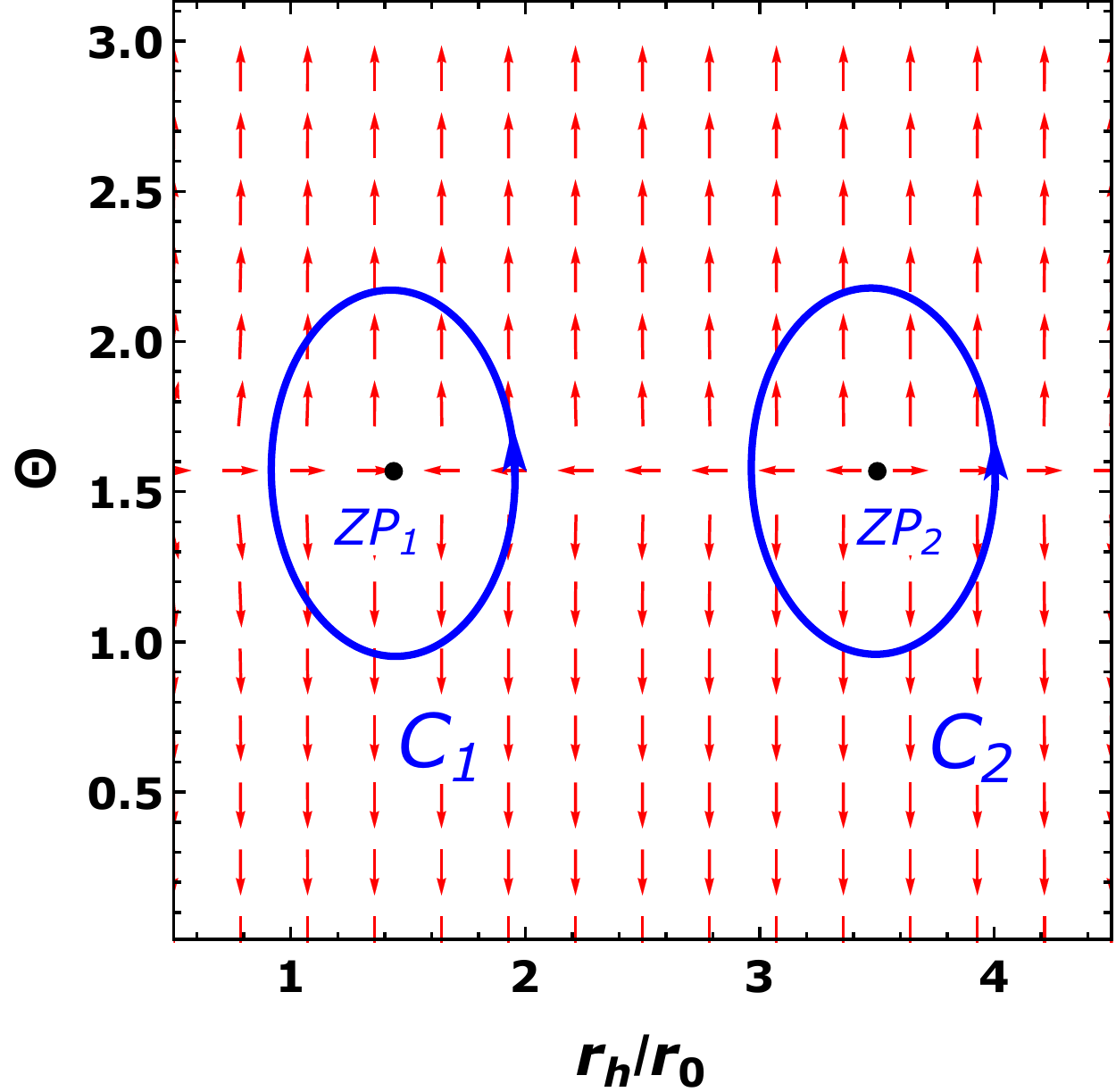}
\caption{The red arrows represent the unit vector field $n$ on a portion of the $r_h-\Theta$
plane with $P=0.002$, $q/r_0=1$, and $\tau/r_0=25$ for the charged BTZ black hole. The zero
points (ZPs) marked with black dots are at $(r_h/r_0, \Theta) = (1.37,\pi/2)$, $(3.63,\pi/2)$
for ZP$_1$ and ZP$_2$, respectively. The blue contours $C_i$ are closed loops surrounding the
zero points. \label{chargedBTZ3d}}
\end{figure}

In Figs. \ref{3dchargedBTZ} and \ref{chargedBTZ3d}, we take $P = 0.002$ and $q/r_0 = 1$ for
the three-dimensional charged BTZ black hole, and plot the zero points of $\phi^{r_h}$ in
the $r_h-\tau$ plane and the unit vector field $n$ with $\tau/r_0 = 25$, respectively. Note
that for the values of $P = 0.002$ and $q/r_0 = 1$, one annihilation point can be found at
$\tau/r_0 = \tau_c/r_0 = 28.03$. Based on the local property of the zero points, one can
get its topological number $W = 0$.

Note that by simply eliminating the charge $q$ in Eq. (\ref{tauA}), the zero point
of the vector field $\phi$ of the static neutral BTZ black hole can be directly expressed as
$\tau = 1/(4Pr_h)$, and its topological number can be easily obtained as $W = 1$, which implies
that the electric charge has a remarkable impact on the topological number of static AdS$_3$
black holes.

\end{document}